\pdfoutput=1
\documentclass[aps,prd,preprint,superscriptaddress,tightenlines,nofootinbib]{revtex4}

\DeclareMathSymbol{\varChi}{\mathord}{letters}{88}
\long\def\symbolfootnote[#1]#2{\begingroup%
\def\thefootnote{\fnsymbol{footnote}}\footnote[#1]{#2}\endgroup}

\usepackage{amsfonts,amsmath,amsthm,amssymb}
\usepackage{mathrsfs}
\usepackage{bm}
\usepackage{color}
\usepackage{epsfig}

\usepackage{dsfont}
\usepackage{mathcomp}

\usepackage{enumitem}

\newcommand{\PRE}[1]{{#1}}   

\newcommand{\beq}{\begin{equation}}
\newcommand{\eeq}{\end{equation}}
\newcommand{\bea}{\begin{flushleft} \begin{eqnarray}}
\newcommand{\eea}{\end{eqnarray}\end{flushleft}}

\newcommand{\postscript}[2]{\setlength{\epsfxsize}{#2\hsize}
   \centerline{\epsfbox{#1}}}
\newcommand{\comment}[1]{}

\newcommand{\n}[1]{$#1$}

\newcommand{\ci}[1]{}

\newcommand{\ba}{\begin{eqnarray}}
\newcommand{\ea}{\end{eqnarray}}
\newcommand{\be}{\begin{equation}}
\newcommand{\ee}{\end{equation}}
\newcommand{\bay}[1]{\left(\begin{array}{#1}}
\newcommand{\eay}{\end{array}\right)}

\def\k{\kappa}                    

\def\m{\mu}
\def\n{\nu}
  
\def\6{\partial}

\def\ed{\end{document}}

\usepackage[usenames,dvipsnames]{xcolor}
\definecolor{orange}{cmyk}{0,0.5,1,0}
\definecolor{rossoCP3}{cmyk}{0,.88,.77,.40}
\definecolor{graa}{rgb}{0.8,0.8,0.8}
\definecolor{blaa}{rgb}{0.2,0.2,0.6}

\begin{document}

\preprint{
\hfil
\begin{minipage}[t]{3in}
\begin{flushright}
\vspace*{-.05in}
MPP-2019-235\\
LMU-ASC 39/19\\
\end{flushright}
\end{minipage}
}

\title{\PRE{\vspace*{-0.5in}} \color{rossoCP3}{ $\bm{H_0}$ tension and the String
    Swampland}}


\author{\bf Luis A. Anchordoqui}

\affiliation{Department of Physics and Astronomy,\\  Lehman College, City University of
  New York, NY 10468, USA
\PRE{\vspace*{.05in}}
}

\affiliation{Department of Physics,\\
 Graduate Center, City University
  of New York,  NY 10016, USA
\PRE{\vspace*{.05in}}
}

\affiliation{Department of Astrophysics,\\
 American Museum of Natural History, NY
 10024, USA
\PRE{\vspace*{.05in}}
}

\author{\bf Ignatios Antoniadis}
\affiliation{Laboratoire de Physique Th\'eorique et Hautes \'Energies - LPTHE\\
Sorbonne Universit\'e, CNRS, 4 Place Jussieu, 75005 Paris, France
\PRE{\vspace*{.05in}}}

\affiliation{Albert Einstein Center, Institute for Theoretical Physics\\
University of Bern, Sidlerstrasse 5, CH-3012 Bern, Switzerland
\PRE{\vspace*{.05in}}}

\author{\bf Dieter\nolinebreak~L\"ust}

\affiliation{Max--Planck--Institut f\"ur Physik, \\ 
 Werner--Heisenberg--Institut,
80805 M\"unchen, Germany
\PRE{\vspace*{.05in}}
}

\affiliation{Arnold Sommerfeld Center for Theoretical Physics 
Ludwig-Maximilians-Universit\"at M\"unchen,
80333 M\"unchen, Germany
\PRE{\vspace{.05in}}
}

\author{\bf Jorge F. Soriano}

\affiliation{Department of Physics and Astronomy,\\  Lehman College, City University of
  New York, NY 10468, USA
\PRE{\vspace*{.05in}}
}

\affiliation{Department of Physics,\\
 Graduate Center, City University
  of New York,  NY 10016, USA
\PRE{\vspace*{.05in}}
}

\author{\bf Tomasz R. Taylor}

\affiliation{Department of Physics,\\
 Northeastern University, Boston, MA 02115, USA 
 \PRE{\vspace*{.05in}}
}

\PRE{\vspace*{.4in}}


\begin{abstract}
\vskip0.1cm
\noindent We realize the Agrawal-Obied-Vafa (AOV) swampland proposal
of fading dark matter by the model of Salam-Sezgin and its string
realization of Cveti\v c-Gibbons-Pope.  The model describes a
compactification of 6-dimensional supergravity with a monopole
background on a 2-sphere. In 4 dimensions, there are 2 scalar fields,
$X$ and $Y $, and the effective potential in the Einstein frame is an
exponential with respect to $Y$ times a quadratic polynomial in the
field $e^{-X}$. When making the volume of the 2-sphere large, namely
for large values of $Y$, there appears a tower of states, which
according to the infinite distance swampland conjecture becomes
exponentially massless. If the standard model fields are confined on Neveu-Schwarz 5-branes the 6-dimensional gauge couplings are independent of the
string dilaton in the string frame, and upon compactification to 4
dimensions the 4-dimensional gauge couplings depend on $X$ (rather
than the dilaton $Y$) which is fixed at the minimum of the
potential. This avoids direct couplings of the dilaton to matter
suppressing extra forces competing with gravity. We show that this set up has
the salient features of the AOV models, and ergo can
potentially ameliorate the tension between
 local distance ladder and cosmic microwave background estimates of
 the Hubble constant $H_0$. Indeed, the tower
of string states that emerges from the rolling of $Y$ constitutes a
portion of the dark matter, and the way in which the $X$ particle and
its Kaluza-Klein excitations evolve over time (refer to as fading dark matter)
is responsible for reducing the $H_0$ tension. Although the AOV proposal
does not fully resolve the tension in $H_0$ measurements, it provides a
dynamical dark energy model of cosmology that satisfies the de Sitter swampland
conjecture. We comment on  a viable solution to overcome the tension
between low- and high-redshift observations within the AOV
background and discuss the implications for the swampland program.
\end{abstract}

\maketitle

\section{Introduction}

Over the past decade or so, and through many experiments, it has
become indisputable that cosmological observations favor an effective
de-Sitter (dS) constant $H$ that nearly saturates the upper bound
given by the present-day value of the Hubble constant,
$H_0$. The $\Lambda$CDM model,
in which the expansion of
the universe today is dominated by the cosmological constant $\Lambda$
and cold dark matter (CDM), is the simplest model that provides a
reasonably good account of all the data. However, various discrepancies have
persisted.  In particular, with the increase in precision of recent
cosmological datasets, measurements of $H_0$ provided by high- and
low-redshift observations started to be in
tension~\cite{Freedman:2017yms}. In the front row, separate
determinations of $H_0$ at low-redshift, including those from Cepheids
and Type-Ia supernovae (SNe), point to
$H_0 = 74.03 \pm 1.42~{\rm km} \, {\rm s}^{-1} \, {\rm
  Mpc}^{-1}$~\cite{Riess:2011yx,Riess:2016jrr,Riess:2018byc,Bonvin:2016crt,Birrer:2018vtm,Riess:2019cxk}. Far from it, when
the sound horizon is calibrated using data from Baryon Acoustic
Oscillations (BAO)  and the all-sky map from the temperature
fluctuations on the cosmic microwave background (CMB), the inferred
 value of the Hubble constant within $\Lambda$CDM  is $H_0 = 67.4 \pm
0.5~{\rm km} \, {\rm s}^{-1} \, {\rm Mpc}^{-1}$~\cite{Hinshaw:2012aka,Ade:2015xua,Aghanim:2018eyx,Verde:2019ivm}. The
discrepancy with the latest SH0ES estimate of $H_0 = 74.03 \pm 1.42~{\rm km} \, {\rm s}^{-1} \, {\rm
  Mpc}^{-1}$~\cite{Riess:2019cxk} is significant at $4.4\sigma$
level~\cite{Verde:2019ivm,Riess:2019cxk}, and systematic effects do
not seem to be responsible for this
inconsistency~\cite{Follin:2017ljs,Dhawan:2017ywl,Shanks:2018rka,Riess:2018kzi,Bengaly:2018xko};
see however~\cite{Rameez}. 

Among the many possible explanations of the $H_0$ tension, those
connecting this discrepancy to the swampland program stand out. The
objective of this program is to extract a set of relatively simple
quantitative requirements for low-energy effective field theories that
admit a UV completion to a consistent theory of quantum gravity~\cite{Vafa:2005ui}. By
now, various swampland conjectures have been
proposed~\cite{ArkaniHamed:2006dz,Ooguri:2006in,Klaewer:2016kiy,Ooguri:2016pdq,Palti:2017elp,Obied:2018sgi,Andriot:2018wzk,Cecotti:2018ufg,Garg:2018reu,Ooguri:2018wrx,Klaewer:2018yxi,Heckman:2019bzm,Lust:2019zwm,Bedroya:2019snp,Kehagias:2019akr,Blumenhagen:2019vgj};
for reviews see~\cite{Brennan:2017rbf,Palti:2019pca}. Of
particular interest here is 
the distance swampland conjecture  that can be
expressed by the following statement: If a scalar field, 
coupled to gravity with  reduced Planck mass
$M_{\rm Pl} = (8 \pi G)^{-1/2}$,
 transverses a trans-Planckian range in the
  moduli space, a tower of string states becomes light exponentially
  with increasing
  distance~\cite{Ooguri:2006in,Klaewer:2016kiy, Ooguri:2018wrx,Grimm:2018ohb,Heidenreich:2018kpg,Laliberte:2019sqc}. The exponentially large number of massless string states saturate the
covariant entropy bound in an accelerating universe~\cite{Bousso:1999xy,Bousso:2006ge}, and force the
scalar field to satisfy the 
so-called de Sitter swampland conjecture~\cite{Ooguri:2018wrx}: The
gradient of the potential $V$ of a canonically normalized scalar field
in a consistent gravity theory must satisfy either the bound,
$M_{\rm Pl} |\nabla V| \geq  c~V $ or must satisfy $M_{\rm Pl}^2 \,
{\rm min}(\nabla_i \nabla_j V)\leq - c' V$,
where $c$ and $c'$ are positive order-one
numbers \cite{Obied:2018sgi,Ooguri:2018wrx}.
 Note that the constraint above  
precludes dS vacua where
$\nabla V = 0$, and therefore rules out $\Lambda$CDM, even when $c \ll 1$~\cite{Agrawal:2018own}.

Studies of dynamical dark energy models that alleviate the $H_0$
tension have been carried out independently of the validity of the
swampland
conjectures~\cite{DiValentino:2016hlg,DiValentino:2017zyq}. One
interesting type of models in this category deals with the scalar
field playing the role of early dark energy, viz. the field could
behave like a cosmological constant at early times (redshifts
$z \agt 3000$) and then dilute away like radiation or faster at later
times~\cite{Poulin:2018cxd,Kaloper:2019lpl,Agrawal:2019lmo,Sakstein:2019fmf}. If
this were the case, the sound horizon at decoupling would be reduced
resulting in a larger $H_0$ value inferred from BAO and CMB
data. However, the CMB-preferred value of  $\sigma_8$ (the rms density fluctuations
within a top-hat radius
of $8 h_0^{-1}~{\rm Mpc}$, with $h_0$ the dimensionless Hubble constant) increases in early dark energy models as
compared to $\Lambda$CDM, increasing the tension with large-scale
structure (LSS) data. More concretely, it is the combination $S_8
= \sigma_8  (\Omega_m/0.3)^{0.5}$ that is constrained by LSS
data, where $\Omega_m$ is the matter density.
The Planck Collaboration reported $S_8 = 0.830 \pm 0.013$~\cite{Aghanim:2018eyx} whereas local measurements find
the smaller values; namely, $S_8^{\rm SZ} = \sigma_8
(\Omega_m/0.27)^{0.3} = 0.78 \pm 0.01$ from Sunyaev-Zeldovich cluster
counts~\cite{Ade:2013lmv}, $S_8 = 0.773^{+0.026}_{-0.020}$ from
DES~\cite{Abbott:2017wau} and $S_8 = 0.745 \pm 0.039$ from
KiDS-450~\cite{Hildebrandt:2016iqg} weak-lensing-surveys. The physical
origin for the increase of $\sigma_8$ in early dark energy models is fairly
straightforward, because the new dark-energy-like component acts to
slightly suppress the growth of perturbations during the period in
which it contributes non-negligibly to the cosmic energy
density. Henceforth, if we want to preserve the fit to the CMB data we must
increase the CDM component to compensate for the suppression in the efficiency of perturbation growth~\cite{Hill:2020osr}.

A second type of interesting models emerges if dark energy and dark
matter interact with each
other~\cite{Salvatelli:2014zta,Valiviita:2015dfa,Abdalla:2014cla,Murgia:2016ccp,Kumar:2017dnp,DiValentino:2017iww,Yang:2018euj,Kumar:2019wfs,DiValentino:2019ffd,DiValentino:2019jae}. The
identification of the infinite tower of string states (following the
swampland distance conjecture) as inhabiting the dark sector
automatically provides a string framework for a concomitant coupling
of the scalar field to the dark
matter~\cite{Agrawal:2019dlm,Vafa:2019evj}. Within this framework
there is a continually reduction of the dark matter mass as the scalar
field rolls in the recent cosmological epoch. Such a reduction of the
dark matter mass is actually compensated by a bigger value of dark
energy density, which becomes visible in the present accelerating
epoch calling for an increase of $H_0$. In this paper we present a
well motivated realization of the cosmological string framework put
forward by Agrawal, Obied, and Vafa (AOV)~\cite{Agrawal:2019dlm}. A
point worth noting at this juncture is that the AOV models do not
fully resolve the tension in $H_0$ measurements, as they can raise the
$\Lambda$CDM predicted value of the Hubble constant only up to
$H_0 = 69.06^{+ 0.66}_{-0.73}~{\rm km} \, {\rm s}^{-1} \, {\rm
  Mpc}^{-1}$~\cite{Agrawal:2019dlm}. Indeed, this maximum value of
$H_0$ is characteristic of all models with late dark energy
modification of the $\Lambda$CDM expansion history. This is because
the local distance ladder calibrates SNe far into the Hubble flow and
if dark matter fading takes place too recently then it would raise
$H_0$ but without actually changing the part of the Hubble diagram
where the tension is inferred. More concretely, by substituting the
SH0ES calibration to the Pantheon SNe dataset, the ability of late times dark energy transitions  to reduce the Hubble tension drops
effectively to $H_0 = 69.17 \pm 1.09~{\rm km} \, {\rm s}^{-1} \, {\rm
  Mpc}^{-1}$~\cite{Benevento:2020fev}. However, the AOV proposal
provides a novel cosmological set up that improves the fit to data compared
to $\Lambda$CDM, while satisfying the dS swampland conjecture. Moreover,
the smaller content of CDM at late times in AOV models as compared to
$\Lambda$CDM yield a slight decrease of $S_8$, which can help reduce
somewhat the tension between the CMB and  LSS datasets.

Our starting point is Salam-Sezgin 6-dimensional supergravity (SUGRA)
model, where a supersymmetric solution of the form
Minkowski$_{4} \times S^2$ is known to exist, with a $U(1)$ monopole
serving as background in the two-sphere~\cite{Salam:1984cj}.  This
model can be lifted to string (and M) theory~\cite{Cvetic:2003xr} and
is asymptotic at large distances to the near-horizon limit of
NS5-branes described by the linear dilaton background which is an
exact string solution~\cite{Antoniadis:1988aa}. Moreover, the
cosmological content of this supergravity model provides a solution of
the field equations that can accommodate both the observed dark energy
density and a fraction of CDM~\cite{Anchordoqui:2007sb}. (Time
dependence in the moduli fields vitiates invariance under
supersymmetry transformations.)  The carrier of the acceleration in
the present dS epoch is a quintessence field slowly rolling down its
exponential potential.  Intrinsic to this model is a second modulus,
which is automatically stabilized and acts as a source of CDM, with a
mass proportional to an exponential function of the quintessence
field. The exponential functional form of the mass spectrum
characterizes the infinite tower of mass states (inherent to the
swampland distance conjecture), which emerges when the quintessence
field moves a distance in field space $\agt {\cal O} (1)$ in Planck
units.

In the proposed cosmological framework,
the standard model (SM) fields are confined to a probe brane and arise from
quantum fluctuations. On the other hand, by computing the quantum fluctuations of the $U(1)$ field associated to the background
configuration it is easily seen that the Kalb-Ramond field generates a
mass term of horizon size~\cite{Anchordoqui:2007sb}. These
``paraphotons'' (denoted herein by $\varUpsilon$) have been redshifting
down since the quantum gravity era without being subject to
reheating. The
presence of any additional relativistic particle species with $g$
degrees of freedom is usually characterized by
\begin{equation}
\Delta N_{\rm eff} \equiv N_{\rm eff} -
N_{\rm eff}^{\rm SM} =  g \ \left(\frac{10.75}{g_*
    (T_{\rm dec})}\right)^{4/3} \times \left\{ \begin{array}{cl} 4/7
                                                &~~{\rm boson} \\
                                                1/2&~~{\rm
                                                     fermion} \end{array}
                                                 \right. \,,
\label{eq:Neff}
\end{equation}
where $N_{\rm eff}$ quantifies the total relativistic ``dark'' energy
density (including the three left-handed SM neutrinos) in units of the
density of a single Weyl neutrino species~\cite{Steigman:1977kc} and
$N_{\rm eff}^{\rm SM} = 3.046$~\cite{Mangano:2005cc}, and where 
$T_{\rm dec}$ is the temperature at which
particle species decouple from the primordial plasma  and the function
$g_*(T_{\rm dec})$ is the number of effective degrees of freedom
(defined as the number of independent states with an additional factor
of 7/8 for fermions) of the SM particle content at the temperature
$T_{\rm dec}$.  Comparing the
106.75 degrees of freedom of the SM with the 10.75 degrees of freedom 
of the primordial plasma before neutrino decoupling it is
straightforward to see that for a massless (real) spin-$0$ scalar,
spin-$\tfrac{1}{2}$ (Weyl) fermion, and massive spin-$1$ vector boson the
contributions to $N_{\rm eff}$ asymptote to specific values of $\Delta
N_{\rm eff} = 0.027$, $0.047$, and $0.080$;
respectively~\cite{Kolb:1990vq}.\footnote{Asymptote here refers to
  relativistic species decoupling just before top quark freeze-out.}  Hence, fluctuations in the Kalb-Ramond field do not influence the primordial abundances of the
nuclides produced at big-bang nucleosynthesis (BBN) as the $\varUpsilon$'s only count
for $\Delta N_{\rm eff} \alt 0.080$ and the 95\% CL limit from a combination of current CMB, BAO, and BBN observations is
$\Delta N_{\rm eff} < 0.214$~\cite{Aghanim:2018eyx}.\footnote{This
  limit combines the helium measurements of~\cite{Aver:2015iza,Peimbert:2016bdg} with the latest
  deuterium abundance measurements of~\cite{Cooke:2017cwo} using the
  the \texttt{PArthENoPE} code~\cite{Pisanti:2007hk} 
considering $d(p,\gamma)^3{\rm He}$ reaction rates from~\cite{Marcucci:2015yla}. Should one instead
use the helium abundance measurement of~\cite{Izotov:2014fga} in place
of~\cite{Aver:2015iza,Peimbert:2016bdg},
 the 95\% CL limit on the equivalent neutrino species shifts,
 $N_{\rm eff} = 3.37 \pm 0.22$, and is in $2.9 \sigma$ tension with the SM value.}

The layout of the paper is as follows. In Sec.~\ref{sec:2} we briefly
describe the geometrical properties of unified dS-Friedmann models
when embedded into Salam-Sezgin 6-dimensional supergravity. In
Sec.~\ref{sec:3} we interpret numerical results from data analysis
that feature estimates for each free parameter in the model.  We show
that the Salam-Sezgin cosmological set up has the
salient features of the generic Agrawal-Obied-Vafa model, and ergo can
potentially ameliorate the tension between
 local distance ladder and cosmic microwave background estimates of
 $H_0$. In Sec.~\ref{sec:4} we comment on a viable solution to overcome the tension
between low- and high-redshift observations within the AOV
background. The paper wraps up with discussion and conclusions
presented in Sec.~\ref{sec:5}. Before proceeding, we note that other ideas
relating cosmological observations to the swampland conjectures have
been presented
in~\cite{Colgain:2018wgk,Heisenberg:2018yae,Akrami:2018ylq,Chiang:2018lqx,Raveri:2018ddi,Elizalde:2018dvw,Brahma:2019kch,Colgain:2019joh,Baldes:2019tkl,Brax:2019rwf,Yang:2018xah,Tosone:2018qei,Achucarro:2018vey,Kehagias:2018uem,Matsui:2018bsy,Ben-Dayan:2018mhe,Kinney:2018nny}.

  \section{Embedding of \lowercase{d}S-Friedmann model into Salam-Sezgin SUGRA}
  
\label{sec:2}

Concentrating on the purely bosonic field content of Salam-Sezgin
6-dimensional 
SUGRA, we can express the bulk action of the system by \beq S \supset \frac{1}{4
  \kappa^2}\int d^6x \sqrt{g_6}\Big[ R_6 - \kappa^2
(\partial_M\sigma)^2 -\kappa^2 e^{\kappa\sigma}F_{MN}^2
-\frac{2g^2}{\kappa^2}e^{-\kappa\sigma}
-\frac{\kappa^2}{3}e^{2\kappa\sigma} G_{MNP}^2 \Big] \,\, ,
\label{ss}
\eeq
where $g_6=\det g_{MN},$ $R_6$ is the Ricci scalar of $g_{MN},$ $\sigma$
is a scalar field,
$F_{MN}=\partial_{[M} A_{N]},$ $G_{MNP}=\partial_{[M}B_{NP]}
+\k A_{[M} F_{NP]}$, $A_N$ is a gauge field, $B_{NP}$ is the
Kalb-Ramond field,  $g$ is the $U(1)$ coupling constant, $\kappa$ the
gravitational coupling constant, and capital Latin indices run from 0 to 5~\cite{Salam:1984cj}. With
redefinition of constants
$G_6\equiv2 \kappa^2$ and $\xi\equiv 4\,g^2$,
and rescaling of $\phi \equiv -\k\sigma$  the action (\ref{ss}) takes the form
\beq
S \supset \frac{1}{2 G_6} \int d^6x\sqrt{g_6}\Big[R_6 -  (\partial_M\phi)^2 -
\frac{\xi}{G_6} e^\phi - \frac{G_6}{2} e^{-\phi} F_{MN}^2 -
\frac{G_6}{6}
e^{-2\phi} G_{MNP}^2\Big] \,\,,
\label{action}
\eeq
where the length dimensions of the
fields are: $[G_6]=L^4,$ $[\xi]=L^2,$ $[\phi]=[g_{MN}^2]=1,$
$[A_M^2]=L^{-4},$ and $[F_{MN}^2]=[G_{MNP}^2]= L^{-6}.$

Note that by rescaling the 6-dimensional metric as $g_{MN}\to
e^{-\phi}g_{MN}$, one finds the action at the string frame where
$\phi$-dependence enters as an overall exponential factor
$e^{-2\phi}$. $\phi$ is then identified with the string dilaton,
defining the string coupling $e^\phi$ and having a tree-level
potential corresponding to a non-critical string with the parameter
$\xi$ determined by the central charge deficit. The latter is induced
by the compactification of the four internal dimensions on a manifold
with non-vanishing curvature. Its sign implies that the internal
curvature is negative, such as the non-compact $H^{(2,2)}\times S^1$
space considered in~\cite{Cvetic:2003xr} to compactify from 10 to 6 dimensions. Its compact analytic
continuation is $S^3\times S^1$, which has an exact (super)-conformal
field theory description, since $S^3$ corresponds to an $SU(2)_k$
Wess-Zumino-Witten model with curvature fixed by the level $k$. The
total internal 6-dimensional space of our model is then $H^{(2,2)} \times S^1
\times S^2$, with the monopole field on $S^2$. The exponential dilaton potential does not allow for static solutions. One solution is the linear dilaton background along a space direction which has an exact string description in terms of a free coordinate with background charge. It corresponds to the near horizon limit of NS5-branes which is holographic dual to a little string theory~\cite{Aharony:1998ub}. In our case of interest, $\xi$ is positive and the solution becomes linear dilaton in the time coordinate with flat metric in the string frame ($\sigma$-model)~\cite{Antoniadis:1988aa}. In the Einstein frame, the scale factor of the metric in FRW coordinates grows linearly with time while the dilaton dependence becomes logarithmic. This exact time dependent `vacuum' solution is the only asymptotic at large times, even in the presence of matter, as we will see later.

We can now carry out a spontaneous compactification from six to four dimensions,
considering the 6-dimensional manifold $M$ of the base spacetime to be
a direct product of 4 Minkowski directions (hereafter denoted by
$M_4$) and the 2-sphere, $\mathbb R^{1,3} \times S^2$. The line
element on $M$ locally is given by
\ba 
ds_6^2= ds_4(t,{\vec x})^2 + e^{2f(t,{\vec x})} \
   r_c^2\ (d\vartheta^2 +\sin^2\vartheta d\varphi^2) \,,
\label{metric}
\ea
where $(t,\ \vec x)$ denotes a local coordinate system in $M_4$, $r_c$ is
the compactification radius, and $f$ is the breathing mode of the
compact space. We assume that the scalar field $\phi$ depends only on the point of $M_4$, i.e., $\phi=\phi(t,\ \vec x)$. We further
assume that the gauge field $A_M$ is excited on $S^2$ and is of the
form
\begin{equation}
A_\vartheta = 0 \quad {\rm and}  \quad
A_\varphi=
b\cos \vartheta\,;
\label{FC}
\end{equation}
this is the monopole configuration detailed
in~\cite{Salam:1984cj}. For the purpose of this work, we will set the
Kalb-Ramond field to its zero background value, $B_{NP} = 0$, and
since the term
$A_{[M} F_{NP]}$ vanishes on $S^2$, we have $G_{MNP} = 0$. The field strength becomes
\begin{equation}
F_{MN}^2=
    2b^2 e^{-4f}/r_c^4  \, .
\label{FS}
\end{equation}
Taking the variation of the gauge field $A_M$ in~(\ref{action})
we obtain the Maxwell equation
\be
\partial_M \Big[ \sqrt{g_4}\sqrt{g_\sigma} e^{2f-\phi} F^{MN}  \Big]=0.
\label{Maxwell}
\ee
It is straightforward to verify that the field strength in~(\ref{FS}) satisfies~(\ref{Maxwell}).

Without loss of generality, the Ricci scalar can be written as\be
R_6 \equiv R[M]=R[M_4]+e^{-2f}R[S^2]-4\Box f-6(\partial_\m f)^2 \,\,, \ee where
$R[M],$ $R[M_4],$ and $ R[S^2] = 2/r_c^2$ denote respectively the Ricci scalars of the
manifolds $M,$ $M_4$ and $S^2$, with Greek indices running from
0 to 3~\cite{Wald:1984rg}. To simplify the notation hereafter $R_4$ and $R_2$ indicate $R[M_4]$
and $R[S^2]$, respectively. The determinant of the metric can be written as
$\sqrt{g_6}=e^{2f}\sqrt{g_4}\sqrt{g_2},$
where $g_4=\det g_{\m\n}$ and $g_2= r_c^4 \sin^2 \vartheta$ is the determinant of the metric
of $S^2$ excluding the factor $e^{2f}$.
We define the gravitational constant in the four dimension as
\be
\frac{1}{G_4}\equiv \frac{M_{\rm Pl}^2}{2} = \frac{1}{2\, G_6}
\int \sqrt{g_2}  \ (d\vartheta \wedge d\varphi) =
\frac{2 \pi r_c^2}{G_6} \,  .
\ee
Thus and so, by using the field configuration given in~(\ref{FC}) the
action in~(\ref{action}) can be recast as
\be
S \!\supset\! \frac{1}{G_4}\! \int\!\! d^4 x \sqrt{g_4} \Big\{ e^{2f}\big[ R_4 +
e^{-2f} R_2 + 2 (\partial_\mu f)^2 - (\partial_\mu \phi)^2 \big]  -
\frac{\xi}{G_6} e^{2f+\phi} - \frac{G_6 b^2}{r_c^4}\, e^{ -2f-\phi} -\frac{G_6}{2}e^{2f-\phi}F_{\mu\nu}^2
\Big\} \,,
\label{action4string}
\ee
where we included the last term that does vanish identically to show what is the 4-dimensional coupling of gauge fields that come from 6 dimensions in the Neveu-Schwarz (NS)
sector. In the spirit of~\cite{Agrawal:2018own}, we now consider a rescaling of the metric of $M_4$ such that
$\hat{g}_{\mu\nu}\equiv e^{2f} g_{\mu\nu}$ and
therefore $\sqrt{\hat{g}_4}=e^{4f} \sqrt{g_4}.$ The preceding metric transformation brings
the model into the Einstein frame, in which the action given
in~(\ref{action4string}) can be rewritten as
\be S \supset \frac{1}{G_4} \int d^4
x \sqrt{\hat{g}_4} \Big[ R[\hat{g}_4] -4 (\partial_\mu f)^2 -
(\partial_\mu \phi)^2 - \frac{\xi}{G_6} e^{-2f+\phi}-\frac{G_6 b^2}{r_c^4}\,
e^{-6f-\phi} + e^{-4f}R_2 -\frac{G_6}{2}e^{2f-\phi}F_{\mu\nu}^2 \Big] \,,
\label{action4}
\ee
and we can use $\hat{g}_4$ in this frame to define a metric which we
use to measure distances in the field space. The effective Lagrangian density in 4 dimensions takes the form
\beq
\mathscr{L} \supset \frac{\sqrt{g}}{G_4}\,\Big[ R- 4 (\partial_\mu f)^2 - (\partial_\mu
\phi)^2 - V(f,\phi)  \Big],
\label{finallagran}
\eeq
with
\beq
V(f,\phi)\equiv \frac{\xi}{G_6}e^{-2f+\phi} + \frac{G_6 b^2}{r_c^4}\,
e^{-6f-\phi} - e^{-4f}R_2 \,\,,
\eeq
where to simplify the notation we have defined: $g\equiv\hat{g}_4$
and $R\equiv R[\hat{g}_4]$.

Next, we define a new orthogonal basis, $X \equiv (\phi+2f)/\sqrt G_4$
and $Y \equiv (\phi-2f)/\sqrt G_4$, so that the kinetic energy terms in
the Lagrangian are both canonical, i.e.,
\beq
\mathscr{L} \supset \sqrt{g}\left[\frac{R}{G_4} -\frac{1}{2}(\partial X)^2
-\frac{1}{2}(\partial Y)^2 - \tilde V(X,Y)  \right],
\label{ll}
\eeq
where the potential $\tilde V(X,Y) \equiv V( f, \phi)/G_4$ can be re-written (after some elementary algebra) as~\cite{Vinet:2005dg}
\beq
\tilde V(X,Y)= \frac{e^{\sqrt{G_4} Y}}{G_4} \left[ \frac{G_6
b^2}{r_c^4}e^{-2\sqrt{G_4} X} - R_2 e^{-\sqrt{G_4} X} +\frac{\xi}{G_6}
\right] \,\, .
\label{potential}
\eeq
 Note that $Y$ corresponds to the 4-dimensional dilaton.
 The equations of motion for the $X$ and $Y$ fields are 
\beq
  \Box  X = \partial_X \tilde V \quad  {\rm  and} \quad \Box  Y =
  \partial_Y \tilde V \, ,
  \label{eqmo}
\eeq
and the Einstein field equations are
\ba
 R_{\mu\nu} - \frac{1}{2}g_{\mu\nu}R & = &
    \frac{G_4}{2} \left[\left(\partial_\mu X \partial_\nu X -\frac{g_{\mu\nu}}{2}\, \partial_\eta X \, \partial^\eta X \right) \right. \nonumber \\
 & + &
\left. \left(
\partial_\mu Y \partial_\nu Y -\frac{g_{\mu\nu}}{2} \,\partial_\eta Y \, \partial^\eta Y \right) -g_{\mu\nu} \tilde V(X,Y)\right] \, +T'_{\mu\nu},
\label{einsteinb}
\ea
where we have added the matter and radiation stress-energy tensor
$T'_{\mu \nu}$, which also contributes to the evolution of the Universe. 
To allow for a dS era we assume that the metric
takes the form
\beq
ds^2= -dt^2 + e^{2h(t)}d\vec{x}^{\,2}  ,
\label{refc}
\eeq
and that $X$ and $Y$ depend only on the time coordinate, i.e., $X=X(t)$ and $Y=Y(t)$. 

Before proceeding, we pause to present our
notation. Throughout, the subindex zero indicates quantities which are
evaluated today. As
usual, we
normalize the Hubble parameter to its value today introducing an
adimensional parameter  $H_0  = 100~h_0~{\rm
  km \, s^{-1} \, Mpc^{-1}}$. Note that the  
function $h(t)$ in the metric measures the evolution of $H$, with
$h(t_0) = h_0$. Now, we can rewrite (\ref{eqmo}) as
\begin{equation}
\ddot X + 3 \dot h\dot X=  - \partial_X \tilde V \quad
{\rm and} \quad
 \ddot Y + 3 \dot h\dot Y= - \partial_Y \tilde V,
\label{eqy}
\end{equation}
and the non-zero independent components
of~(\ref{einsteinb}) are
\begin{equation}
\dot h^2 = \frac{G_4}{6}\left[\frac{1}{2}(\dot X^2 + \dot Y^2) + \tilde V(X,Y)\right]+\frac{\rho'}{3}
\label{hubble}
\end{equation}
and
\begin{equation}
 2 \ddot h+3\dot h^2 =  \frac{G_4}{2}\left[-\frac{1}{2}( \dot X^2 +  \dot Y^2) + \tilde V(X,Y)\right]-p'\, ,
 \label{pressure}
\end{equation}
where $p$ and $\rho'$ are the pressure and energy density contained in $T'_{\mu\nu}$.

The terms in the square brackets in~(\ref{potential}) take the
form of a quadratic function of $e^{-\sqrt{G_4}\,X}.$ This function
has a global minimum at $ e^{-\sqrt{G_4}\, X_0} = R_2\, r_c^4/(2\,
G_6\,b^2)$, and so we expand~(\ref{potential}) around the minimum,
\begin{equation}
\tilde V(X,Y) =  \frac{e^{\sqrt{G_4}\, Y}}{G_4} \,  \left[
    {\cal K} + \frac{\overline{M_X}^2}{2}  (X-X_0)^2 +
    {\cal O} \Big((X- X_0)^3 \Big)\right] \,,
\label{minBUF}
\end{equation}
where
\begin{equation}
\overline{M_X} \equiv \frac{1}{\sqrt{\pi}\,\,\, b r_c}
\label{mxbar}
\end{equation}
and
\begin{equation}
{\cal K} \equiv \frac{M_{\rm Pl}^2}{4 \pi r_c^2 b^2} (b^2 \xi - 1) \,\, .
\label{calk}
\end{equation}
Obviously the scalar field $X$ is stabilized around its minimum $X_0$. Its physical mass is $Y$-dependent,
\begin{equation}
M_X (Y)= \frac{e^{\sqrt{G_4}\, Y/2}}{\sqrt{G_4}}\ \overline{M_X}\, ,
\label{mphys}
\end{equation}
and characterizes the mass scale of the tower of string states, which 
according to the infinite distance conjecture becomes exponentially massless~\cite{Ooguri:2006in,Klaewer:2016kiy, Ooguri:2018wrx,Grimm:2018ohb,Heidenreich:2018kpg,Laliberte:2019sqc}. Indeed, as $Y$ runs to large and negative values the 4-dimensional Planck mass grows exponentially as $M_{\rm Pl}\sim e^{-Y}$ in string units, and thus string excitations become exponentially light in Planck units. Note though that these states cannot play the role of dark matter since part of the string modes carry also SM gauge charges. The $X$ particle on the other hand can play the role of fading dark matter, as we show in the next section.

In the absence of matter and radiation described by the stress tensor $T'_{\mu\nu}$, the equations of motion \eqref{eqy}-\eqref{pressure} have no dS or inflationary solution. As we mentioned above, there is an exact string solution with both functions $h$ and $Y$ logarithmic in time describing a linearly expanding universe, which corresponds in the string frame to the well know linear dilaton and flat metric background. This requires the parameter $\cal K$ in eq.~\eqref{minBUF} to be positive. As we will see later, this solution becomes asymptotic at large times in the presence of matter and radiation. 
Moreover, there is a period in time of approximate exponential expansion.
The dS (vacuum) potential energy density is given by
\begin{equation}
V_Y = \frac{e^{\sqrt{G_4}\, Y}}{G_4}\ {\cal K} \, .
\label{rowvac}
\end{equation}
Now, the requirements for preserving a fraction of supersymmetry (SUSY) in
spherical compactifications to four dimension imply $b^2 \xi = 1$,
corresponding to winding number $n= \pm 1$ for the monopole
configuration~\cite{Salam:1984cj}. From (\ref{calk}) and
(\ref{rowvac}) it follows that the condition for the potential to show
a dS rather than an AdS or Minkowski phase is $\xi b^2 >
1$. Therefore, we conclude that a ($Y$-dependent) dS background can be obtained only
through SUSY breaking; see Appendix for details.

We finish this section with a comment on possible SM embeddings. In principle, excitations of the electromagnetic field would seemingly induce variation in the electromagnetic fine structure constant, as well as a violation of the equivalence principle through a long range coupling of the dilaton to the electromagnetic component of the stress tensor~\cite{Carroll:1998zi}. A similar variation would be induced in the QCD gauge coupling and thus in the hadron masses. Although a preliminary analysis seems to indicate that such variations may still be compatible with experimental limits because the resulting range of variation of the quintessence field is about 2.5 Planck units (see next section), a very light dilaton would also mediate extra forces at short and larger distances~\cite{Taylor:1988nw} which are excluded in particular by microgravity experiments~\cite{Adelberger:2013jwa}. A possible way out would be to confine the SM fields on NS5-branes~\cite{Antoniadis:2001sw}. The 6-dimensional gauge couplings are then independent of the string dilaton in the string frame and thus come with a factor $e^\phi$ instead of $e^{-\phi}$ in the Einstein frame, see eq.~\eqref{action}. It follows that upon compactification to four dimensions, gauge kinetic terms couple to $e^{2f+\phi}$, see eq.~\eqref{action4}, and thus the 4-dimensional gauge couplings depend on the scalar $X$ (instead of the dilaton $Y$) which is fixed at the minimum of the potential, and SM couplings do not vary. Moreover, one avoids direct couplings of the dilaton to matter suppressing extra forces competing with gravity.

\section{Reducing the $\bm{H_0}$ tension with fading dark matter}
\label{sec:3}

We now turn to investigate the cosmological implications of the
Salam-Sezgin model, by accommodating recent cosmological observations,
while seeking to diminish the tension between low- and high-redshift
measurements. To do so, we adopt the best fit value of $H_0 = 69.06^{+0.66}_{-0.73}~{\rm km} \, {\rm s}^{-1} \, {\rm
  Mpc}^{-1}$ in the
AOV study~\cite{Agrawal:2019dlm} and analyze the dependence of the quantities relevant to cosmology on the model parameters.

The total energy density of the Universe,
$\rho=\sum_i\rho_i$,
drives the evolution of the Hubble parameter $H$,
where $i=\{ X, Y, {\cal X}, b, r\}$ accounts for the $X$ and $Y$ fields, for
other types of dark matter ${\cal X}$, and for the usual SM components of
baryonic matter $b$
and radiation $r$.  For a spatially flat Universe, $H^2=\rho/3$, where we
have adopted reduced Planck units, i.e., $M_{\rm Pl}=1$ and $G_4=\sqrt{2}$. For convenience, herein we consider the evolution in 
$u\equiv-\ln(1+z)$ rather than $t$, where $z$ is the redshift
parameter. With this in mind, we express the evolution of the matter and radiation components as 
\begin{subequations}
\begin{equation}
\rho_{b}  =\rho_{b,0} \ e^{-3u},
\end{equation}
\begin{equation}
  \rho_{\cal X} = \rho_{{\cal X},0}  \ e^{-3u},
\end{equation}
and
\begin{equation}
\rho_{r}=\rho_{r,0} \ e^{-4u} \ f(u),
\end{equation}
\end{subequations}
where we remind the reader that the subindex zero indicates quantities which are evaluated today, and
$f(u)$ parametrizes the $u$ dependent number of radiation degrees of
freedom.  For the sake of interpolating the various thresholds
appearing prior to recombination (among others, QCD and electroweak),
we adopt a convenient phenomenological form derived
elsewhere~$f(u)=e^{-u/15}$~\cite{Anchordoqui:2003ij}. To simplify
notation we also conveniently define $\rho_* = \rho_b + \rho_{\cal X}$. A
point worth noting at this juncture is that the leading term in the
expansion of the potential $\tilde V$ around the local minimum $X_0$ is quadratic, and
therefore the coherent $X$-field energy behaves like non-relativistic
dark 
matter~\cite{Turner:1983he}.  Thus, the $X$ pressureless dark matter and ${\cal X}$ add
up to the CDM of our model. All in all,
the number density of the field $X$ evolves like that of a matter
term (i.e., proportional to $e^{-3u}$), while its mass evolves with $Y$ according to
(\ref{mphys}). Therefore, as in the AOV scheme~\cite{Agrawal:2019dlm}, we have
\begin{equation}
\rho_X=M_X \ n_X=\rho_{X,0} \ \exp\left(\frac{Y-Y_0}{\sqrt2}-3u\right)= A
\exp\left(\frac{Y}{\sqrt2}-3u\right) \, .
\end{equation}
Finally, the energy density for $Y$ is found to be 
\begin{equation}
\rho_Y=\frac12 H^2 Y'^2+V_Y.\label{eq:j4}
\end{equation}

Now, making use of the preceding formulae, we can give an explicit
expression for the evolution of the Hubble parameter:
\begin{equation}
H^2=\frac{\rho_\mathrm s}{3-Y'^2/2},\label{eq:fried}
\end{equation}
where $\rho_\mathrm s=\rho_*+\rho_r+V_{\rm eff}$ stands for the
\textit{steady-state} energy density in moduli space, in the sense
that the field $Y$ is not evolving ($Y'=0$), with $V_{\rm eff} \equiv
V_Y + \rho_X$. These definitions allow us to rewrite the evolution equation (\ref{eqy}) for $Y$ as
\begin{equation}
\frac{Y''}{1 - \frac16 {Y'}^2} \,\, + 3 \,Y' + \frac{\frac12 Y'\partial_u\rho_\mathrm s+3\,\partial_Y V_{\rm eff}}{\rho_\mathrm s}=0.\label{eq:Yfinal}
\end{equation}
Next, to simplify the numerical solution to the last equation, we introduce the parameters
\begin{subequations}
\begin{equation}
\alpha\equiv\frac{V_0}{\rho_{*,0}},\label{eq:j6a}
\end{equation}
\begin{equation}
\beta\equiv\frac{\rho_{r,0}}{\rho_{*,0}},\label{eq:j6b}
\end{equation}
and
\begin{equation}
\gamma\equiv\frac{A}{\rho_{*,0}},\label{eq:j6c}
\end{equation}
\label{eq:abc}
\end{subequations}
where $V_0\equiv \left.V_Y\right|_{Y=0}=\mathcal K/\sqrt2$.
Further definition of $\rho_{\rm
  s}\equiv\rho_{*,0}\,\overline \rho_\mathrm s$ and $
V_\mathrm{eff}\equiv \rho_{*,0}\, \overline V_\mathrm{eff}$, which depend only on
the parameters introduced in (\ref{eq:abc}), makes explicit the
dependence of the solution to (\ref{eq:Yfinal}) on just  $\alpha$,
$\beta$, and $\gamma$. Following~\cite{Anchordoqui:2007sb}, we take as initial conditions
$Y(-30) = 0$ and $Y'(-30) = 0.08$, which are in accordance to equipartition arguments~\cite{Steinhardt:1999nw,LopesFranca:2002ek}.

In order to understand to which extent this model can represent cosmological data, we introduce the density parameters $\Omega_i=\rho_i/3H^2$ and the equation of state for the field $Y$:
\begin{equation}
w_Y\equiv\frac{p_Y}{\rho_Y}=\frac{\frac12H^2Y'^2-V_Y}{\frac12H^2Y'^2+V_Y}.\label{eq:wdef}
\end{equation}
At this stage, it is worthwhile to note that although the solution for
$Y$ only depends on $\alpha$, $\beta$ and $\gamma$, the cosmological
quantities depend on additional parameters. For instance, the use of
(\ref{eq:fried}) and (\ref{eq:wdef}) requires the introduction of
$\rho_{m,0} = \rho_{*,0} + \rho_{X,0}$ and $h_0$ as additional
parameters. This amounts to a total of five free parameters in this
model. For future convenience, they are chosen to be $h_0$,
$\Omega_{m,0}$, $\Omega_{r,0}$, $a\equiv A/(3H_0^2)$ and $v_0\equiv
V_0/(3H_0^2)$. These parameters are constrained by five
conditions. One is the use of (\ref{eq:fried}) as an internal
consistency condition on the total energy density. Four additional
constraints will come as an attempt to reproduce experimental data
with this model. In particular, we will fix $h_0$ to an experimental
value $\tilde h_0$, and subsequently fixing the radiation content of
the universe, since this model does not provide any mechanism to modify it, with the additional constraint
\begin{equation}
\Omega_{r,0}=\tilde\Omega_{r,0}\equiv \frac{\left.\Omega_{r,0}h_0^2\right|_{\rm exp}}{{\tilde h_0}^2}.\label{eq:cons-r}
\end{equation}
The total matter content of our model is similarly adjusted to an
experimental value and is given by
\begin{equation}
\Omega_{m,0} \equiv \Omega_{*,0}
+\Omega_{X,0}=\tilde\Omega_{m,0}=\frac{\left.\Omega_{
      m,0}h_0^2\right|_{\rm exp}}{\tilde h_0^2} \, . \label{eq:cons-M}
\end{equation}
Before we go any further, we clarify that 
a tilde on top of a given parameter
of the model, 
 identifies its direct experimental measurement, and 
when the measured quantity is a product of two model parameters then we
adopt the subindex exp to indicate the experimental
measurement. Finally, the equation of state for $Y$ today is fixed to
the value of the dark energy equation of state $w_{Y,0}=\tilde
w_{Y,0}$. In our calculations we take 
$\tilde w_{Y,0} = -0.80^{+0.09}_{-0.11}$, as derived from a combination
of multiple observational probes in the Dark Energy Survey (DES)
supernovae program (including
207 type Ia supernovae light curves,
the BAO feature, weak gravitational lensing, and galaxy clustered, but
independent of CMB measurements)~\cite{Abbott:2018wzc}. This value of
$\tilde w_{Y,0}$ is consistent at the $1\sigma$ level with the one
derived from a combination of DES data and CMB measurements~\cite{Abbott:2018wog}. 

Making use of (\ref{eq:wdef}) and (\ref{eq:j6a}), we can rewrite the constraint on the equation of state as
\begin{equation}
\tilde w_{Y,0}=\frac{\frac16Y_0'^2-v_0 e^{\sqrt2 Y_0}}{\frac16Y_0'^2+v_0 e^{\sqrt2 Y_0}}.\label{eq:w1}
\end{equation}
Making use of (\ref{eq:fried}) at $u=0$ together with (\ref{eq:cons-r}) and (\ref{eq:cons-M}) we arrive at 
\begin{equation}
\frac16Y_0'^2=1-\tilde\Omega_{r,0}-\tilde\Omega_{m,0}-v_0 e^{\sqrt2 Y_0},\label{eq:yprime}
\end{equation}
which can be substituted into (\ref{eq:w1}) to find the constraint
\begin{equation}
v_0 e^{\sqrt 2Y_0}=c_- \, .\label{eq:cons-v0}
\end{equation}
Moreover, this result can be substituted back
into (\ref{eq:yprime}) to find a second constraint: ${Y'_0}^2=6c_+$.
We have defined the experimentally determined constants
\begin{equation}
c_\pm\equiv \frac{1\pm\tilde
  w_{Y,0}}{2}(1-\tilde\Omega_{m,0}-\tilde\Omega_{r,0}) \, .
\end{equation}
The third independent constraint between the still free parameters
$\Omega_{*,0}$, $a$, and $v_0$ can be found as a result of
(\ref{eq:cons-v0}) and  $\Omega_{X,0}=a e^{Y_0/\sqrt2}$, and is given by  
\begin{equation}
v_0=c_-\left(\frac{a}{\tilde\Omega_{m,0}-\Omega_{*,0}}\right)^2=c_-\left(\frac{a}{\Omega_{X,0}}\right)^2,\label{eq:v0}  
\end{equation}
unless $a=\Omega_{X,0}=0$. Under this condition, $v_0$ can be determined from $a$ and $\Omega_{X,0}$, in which case the full solution comes from the solution to the system
\begin{subequations}
\begin{equation}
Y_0=\sqrt2\ln \left(\frac{\Omega_{X,0}}{a}\right),\label{eq:sys1}
\end{equation}
\begin{equation}
{Y'_0}^2=6c_+.\label{eq:sys2}
\end{equation}\label{eq:sys}
\end{subequations}
It must be noted that both $Y_0$ and ${Y'_0}^2$ are functions of
$\Omega_{X,0}$ and $a$ through their dependence on the parameters
$\alpha$, $\beta$ and $\gamma$ from (\ref{eq:abc}). Solving separately
(\ref{eq:sys1}) and (\ref{eq:sys2}) we can obtain two solutions
$\Omega_{X,0}^{(1)}(a)$ and $\Omega_{X,0}^{(2)}(a)$ respectively. A
common solution exists if there is some $a$ such that
$\Omega_{X,0}^{(1)}(a)=\Omega_{X,0}^{(2)}(a)$. In the case that $a=\Omega_{X,0}=0$ and $\Omega_{*,0}=\Omega_{m,0}$, the remaining parameter $v_0$ cannot be determined through (\ref{eq:v0}), and its values $v_0^{(1)}$ and $v_0^{(2)}$ will come from the solutions to (\ref{eq:cons-v0}) and (\ref{eq:sys2}), respectively,  expressing $Y_0$ and $Y'_0$ as functions of $v_0$. 

In the following we will consider the matter and radiation parameters
as given by the Particle Data Group, 
$\left.\Omega_{b,0}h_0^2\right|_{\rm exp}= 0.02226(23)$,
$\left.\Omega_{{\rm CDM},0}h_0^2\right|_{\rm exp}= 0.1186(20)$,  and
$\left.\Omega_{r,0}h_0^2\right|_{\rm exp}= 2.473 \times 10^{-5}
(T_{\gamma,0}/2.7255)^4$, where $T_{\gamma, 0}$ is the temperature of
the relic photons~\cite{Tanabashi:2018oca}. The existence of
solutions to (\ref{eq:sys}) is conditioned by the values of $\tilde
h_0$ and $\tilde w_{Y,0}$ through the constants $c_\pm$. For example,
for $(\tilde h_0,\tilde w_{Y,0})=(0.71,-0.62)$, there exists a
solution for $(\Omega_{X,0},a)\approx(0.019,0.107)$ but there is no
solution for $(\tilde h_0,\tilde w_{Y,0})=(0.71,-1)$.  A systematic analysis of the $(\tilde h_0,\tilde w_{Y,0})$ parameter space is necessary to study the potential of this model.

\begin{figure}[tb] 
    \postscript{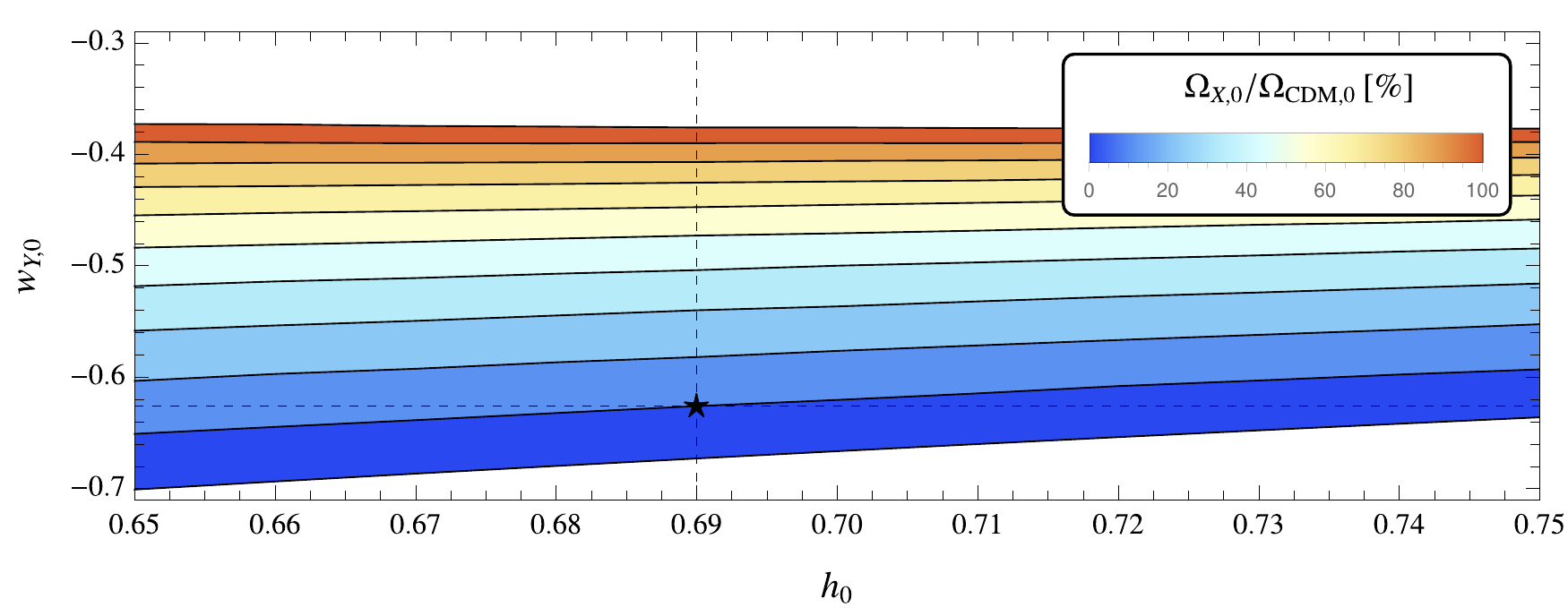}{0.99} 
    \caption{Allowed region in the $(h_0,w_{Y,0})$ parameter space, in terms of the
 ratio of present energy densities for the $X$-field and the total
 CDM. The star indicates the preferred $h_0$ value in the AOV analysis~\cite{Agrawal:2019dlm}. 
\label{fig:parspace}}
\end{figure}

For large values of $a$, it can be seen that $\Omega_{X,0}^{(2)}$ is consistently larger than $\Omega_{X,0}^{(1)}$, in a wide region of the $(\tilde h_0,\tilde w_{Y,0})$ parameter space. This can be used to study the existence of solutions. As $\Omega_{X,0}$ and $a$ go to zero simultaneously, they do it as
 \begin{equation}
 \Omega_{X,0}=\sqrt\frac{c_-}{v_0}\,a,
 \end{equation}
 as follows from (\ref{eq:v0}). To ensure consistency with the solutions at $a=\Omega_{X,0}=0$, each function $\Omega_{X,0}^{(i)}$ must have a different slope
 \begin{equation}
 \Omega_{X,0}^{(i)}=\sqrt\frac{c_-}{v_0^{(i)}}\,a.
 \end{equation}
Using this, if $v_0^{(2)}>v_0^{(1)}$, both curves must cross,
guaranteeing the existence of a solution. The limiting condition
$v_0^{(2)}=v_0^{(1)}$, which determines the existence of a solution
with $a=0$, separates both regions in the $(\tilde h_0,\tilde
w_{Y,0})$ parameter space. In Fig.~\ref{fig:parspace} we show this
limiting condition together with several solutions for
$a\neq0$. The best fit value of~\cite{Agrawal:2019dlm}, $h_0 = 0.69$, is indicated by a star.  We
note that models with $\Omega_{X,0}/\Omega_{\rm CDM,0} \agt 40\%$ are in $3\sigma$ tension with current
determinations of $w_{Y,0}$.

We now take the best fit solution derived in~\cite{Agrawal:2019dlm} as
the experimental value of $H_0$ and check for consistency of the
relevant cosmological parameters. For $h_0=0.69$ and
$\Omega_{X,0}/\Omega_{\rm CDM,0} =0.1$, we obtain $w_{Y,0} = -0.63$, which gives $a
= 0.178$,  $v_0= 35.3$, and  $V_0 = 3H_0^2 v_0 = 3.87 \times
10^{-119}$ in reduced Planck units.

\begin{figure}[tpb]
\begin{minipage}[t]{0.49\textwidth}
\postscript{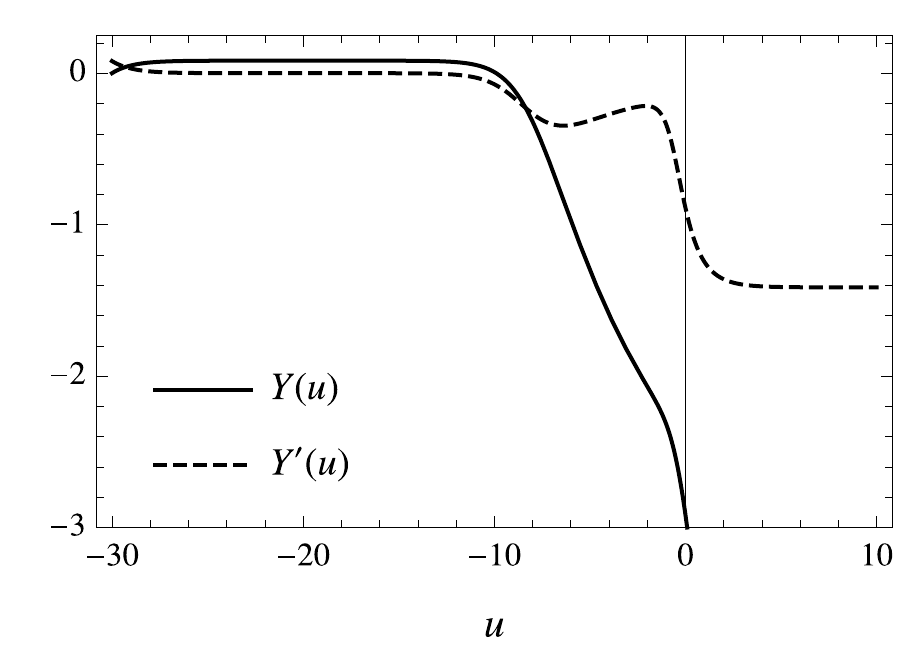}{0.99}
\end{minipage}
\begin{minipage}[t]{0.49\textwidth}
\postscript{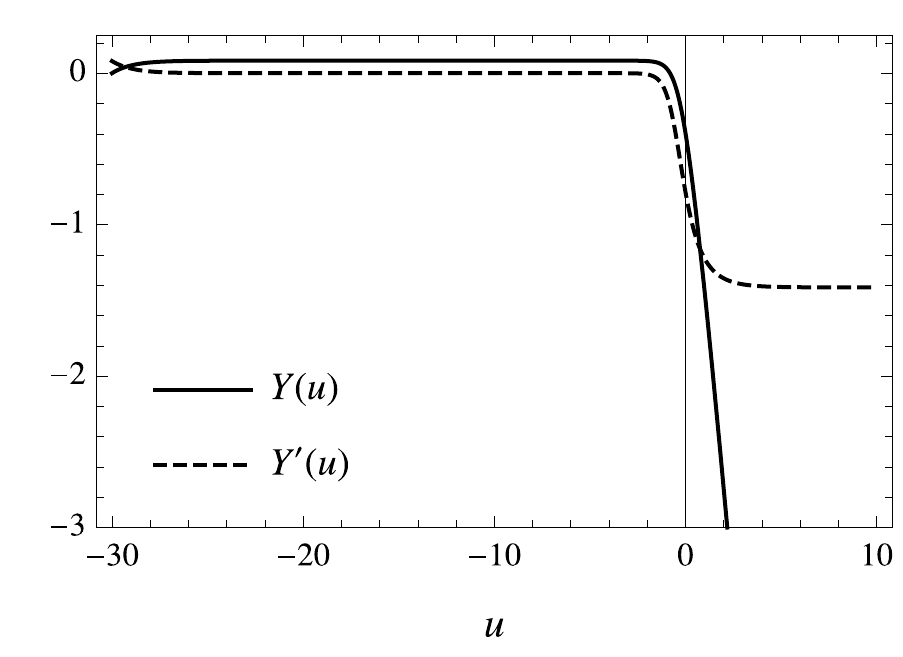}{0.99}
\end{minipage}
\caption{Evolution of $Y(u)$ and $Y'(u)$, for $a=0.178$ (left) and
  $a=0$ (right).  \label{fig:1}}
\end{figure}

\begin{figure}[tpb]
\begin{minipage}[t]{0.49\textwidth}
\postscript{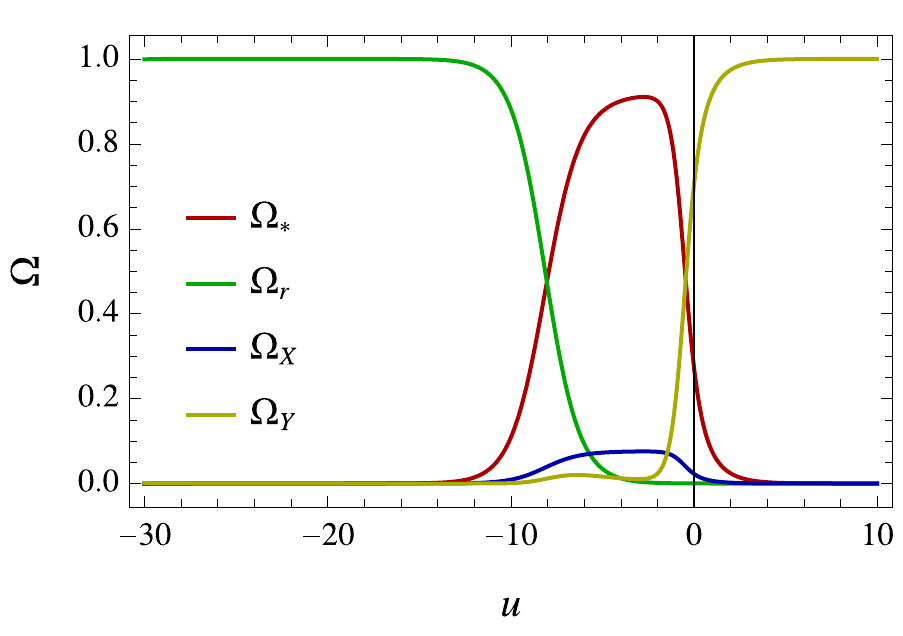}{0.99}
\end{minipage}
\begin{minipage}[t]{0.49\textwidth}
\postscript{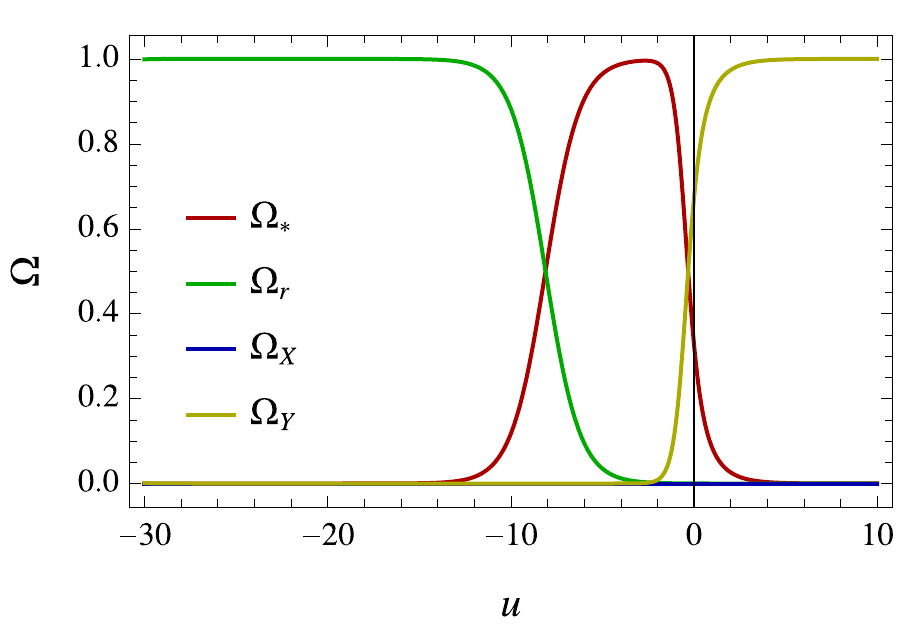}{0.99}
\end{minipage}
    \caption{Evolution of the density parameters $\Omega_r$,
      $\Omega_*$, $\Omega_X$, and $\Omega_Y$. We have taken $a=0.178$ (left) and
  $a=0$ (right). \label{fig:2}}
\end{figure}

\begin{figure}[tpb]
\begin{minipage}[t]{0.49\textwidth}
\postscript{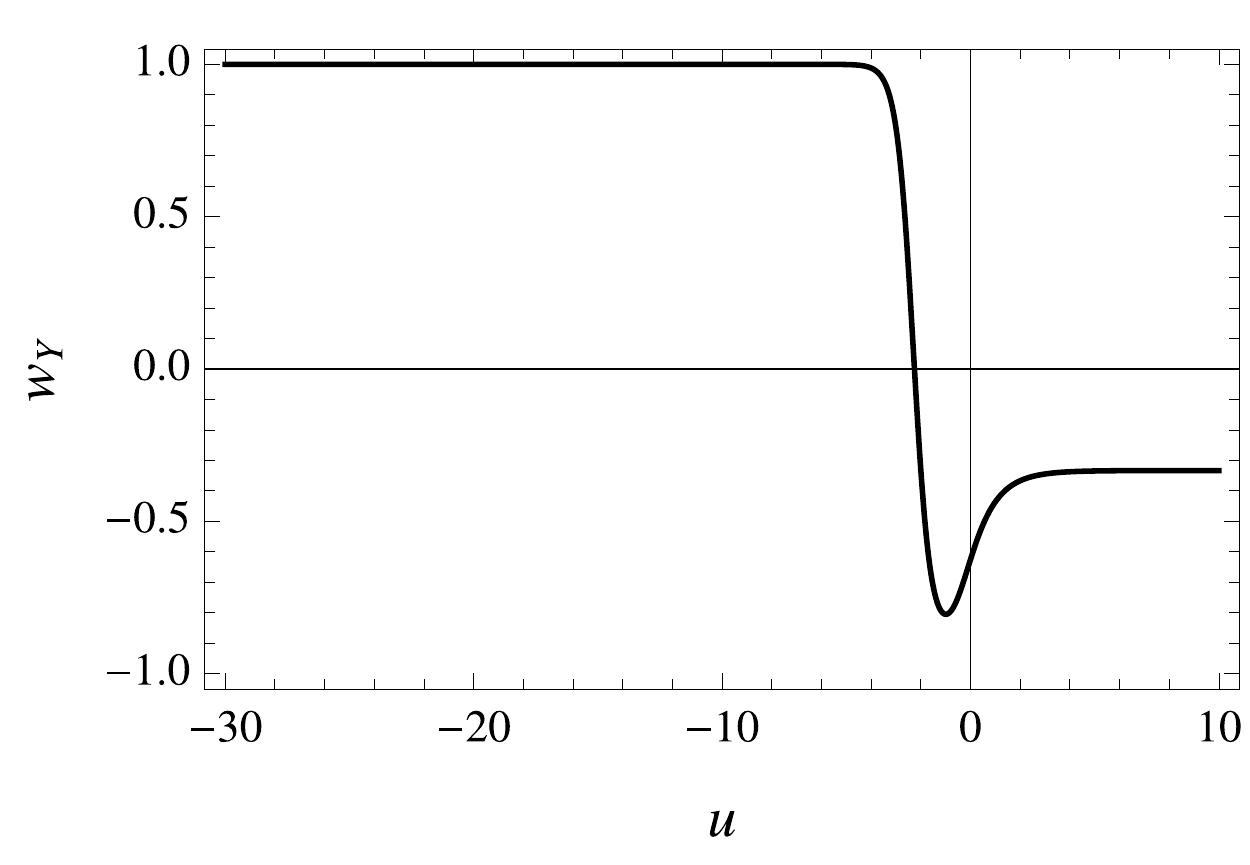}{0.99}
\end{minipage}
\begin{minipage}[t]{0.49\textwidth}
\postscript{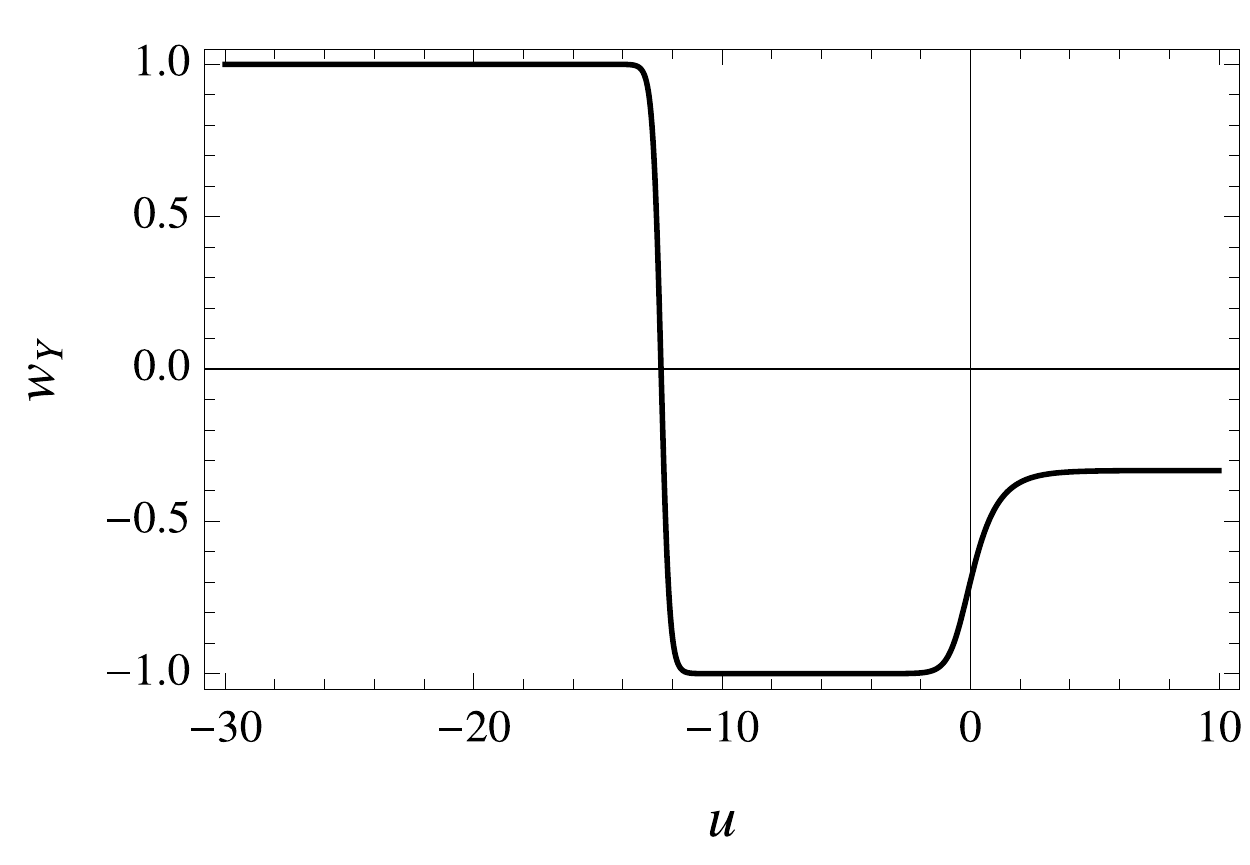}{0.99}
\end{minipage}
    \caption{Evolution of the equation-of-state parameter for dark
      energy $w_Y$, for $a=0.178$ (left) and
  $a=0$ (right). \label{fig:3}}
  \end{figure}

  \begin{figure}[tpb]
\begin{minipage}[t]{0.49\textwidth}
\postscript{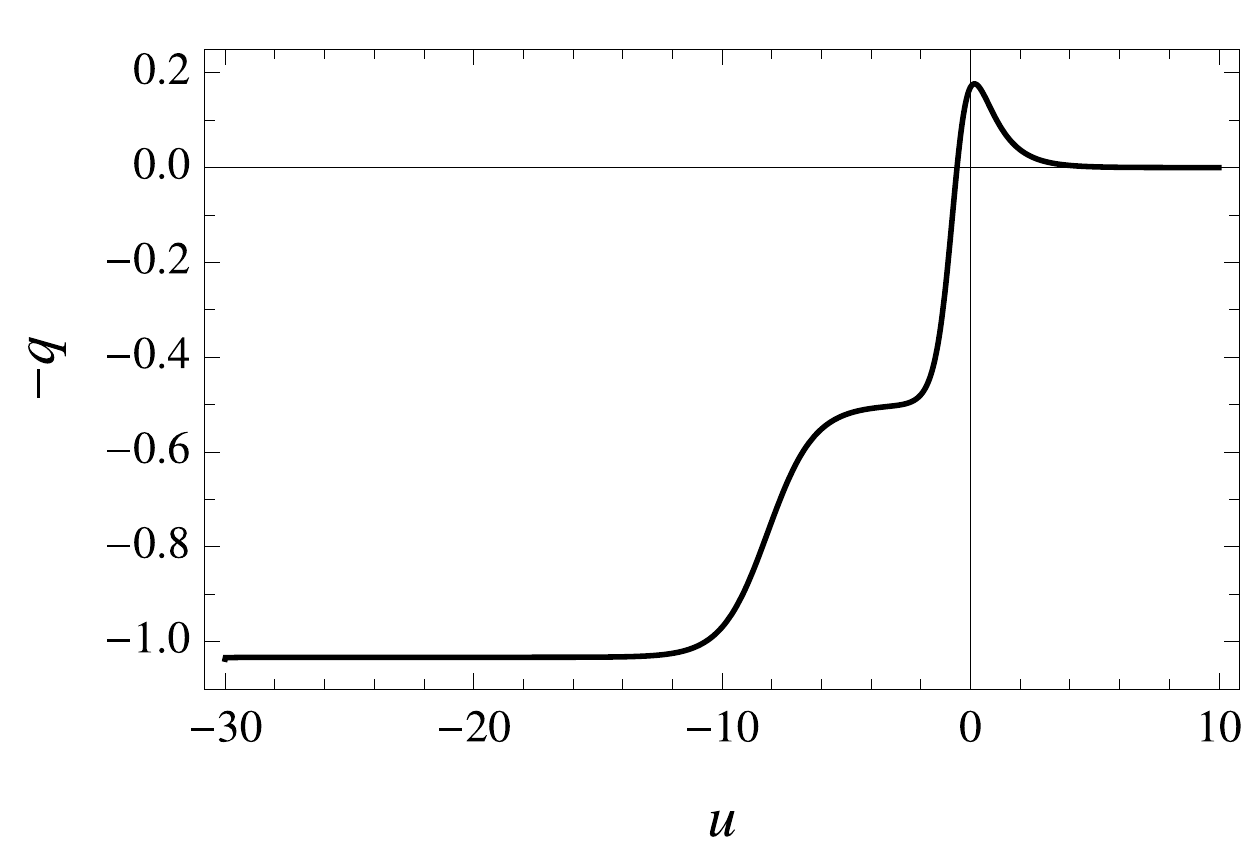}{0.99}
\end{minipage}
\begin{minipage}[t]{0.49\textwidth}
\postscript{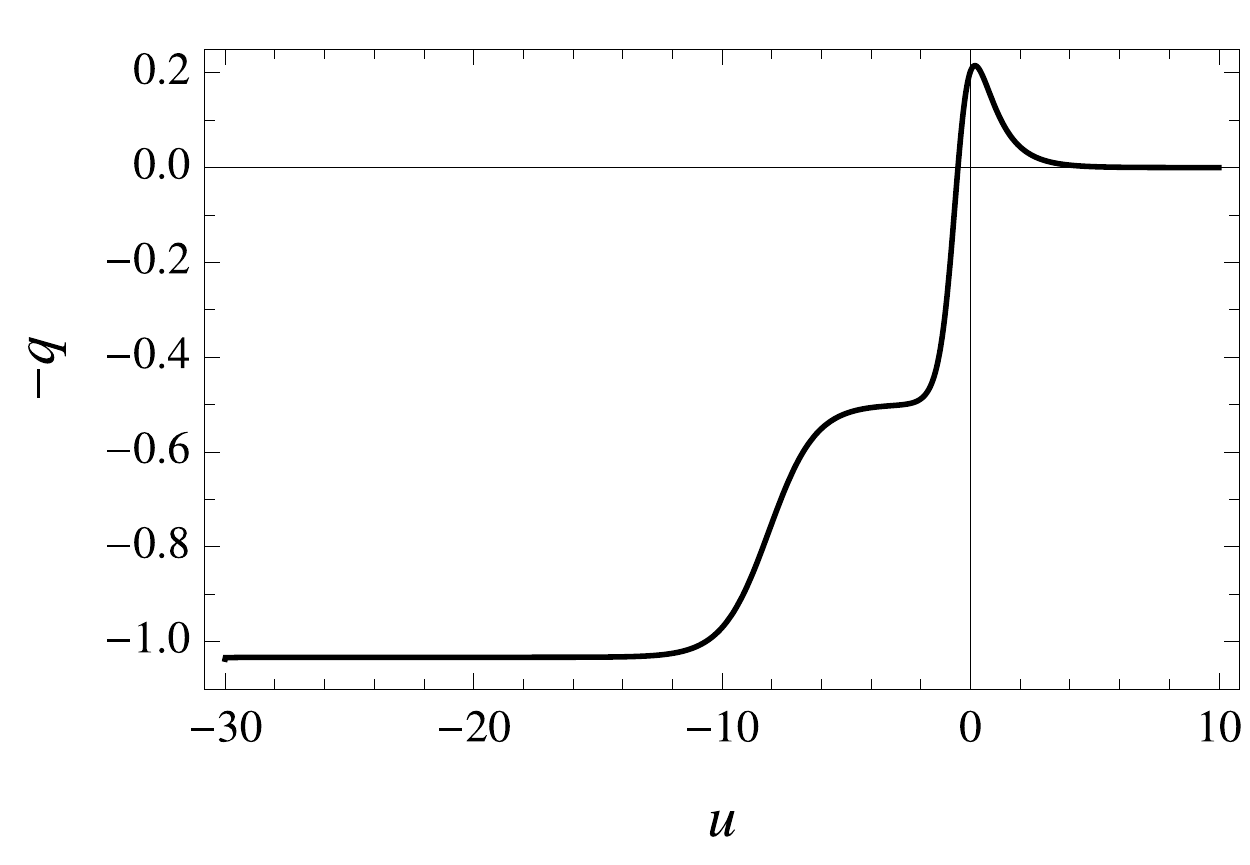}{0.99}
\end{minipage}
    \caption{Evolution of the acceleration parameter $-q$, for $a=0.178$ (left) and
  $a=0$ (right), showing the existence of an accelerated phase that asymptotically approaches a constant velocity expansion in the future. \label{fig:q}}
  \end{figure}

\begin{figure}[tpb]
\begin{minipage}[t]{0.49\textwidth}
\postscript{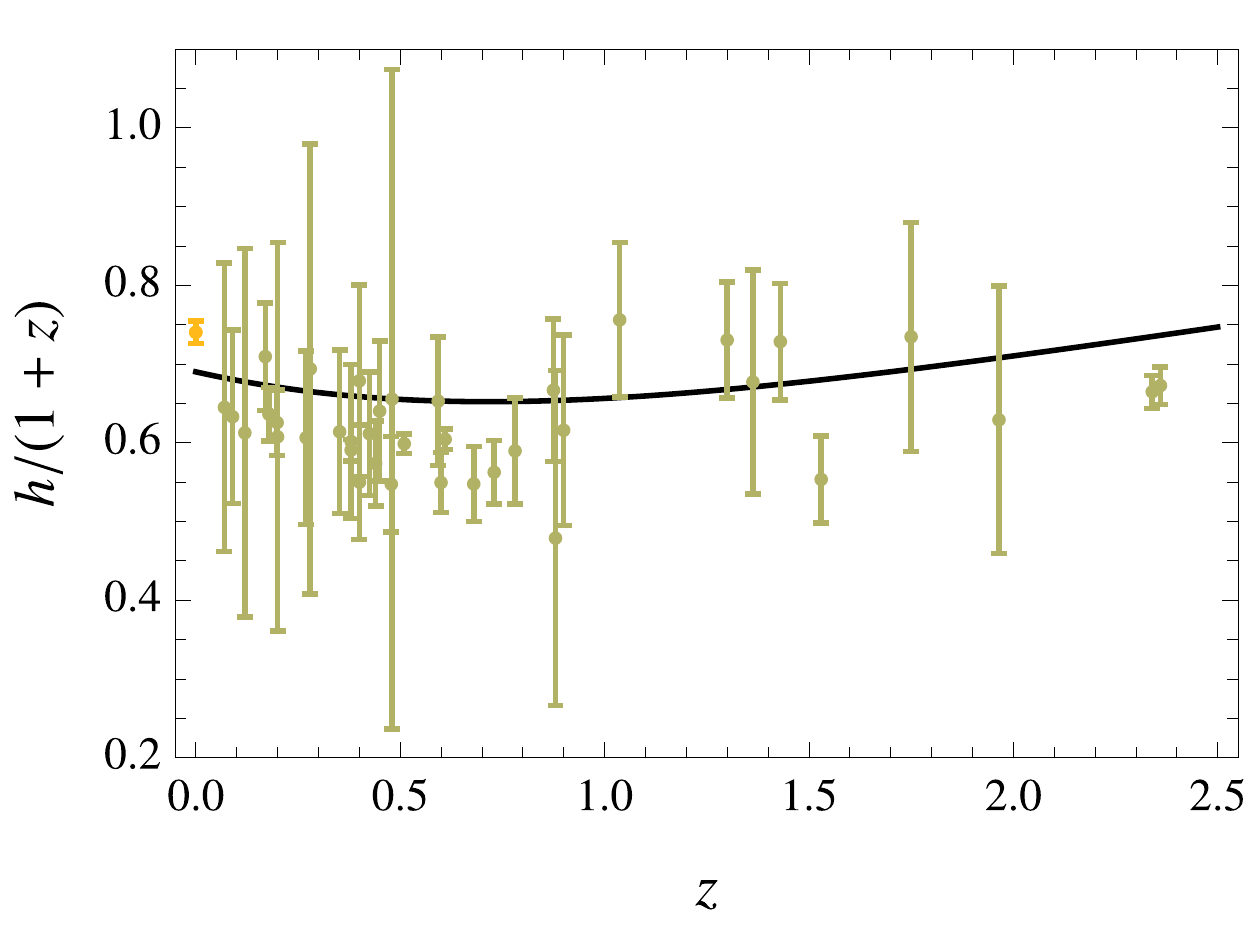}{0.99}
\end{minipage}
\begin{minipage}[t]{0.49\textwidth}
\postscript{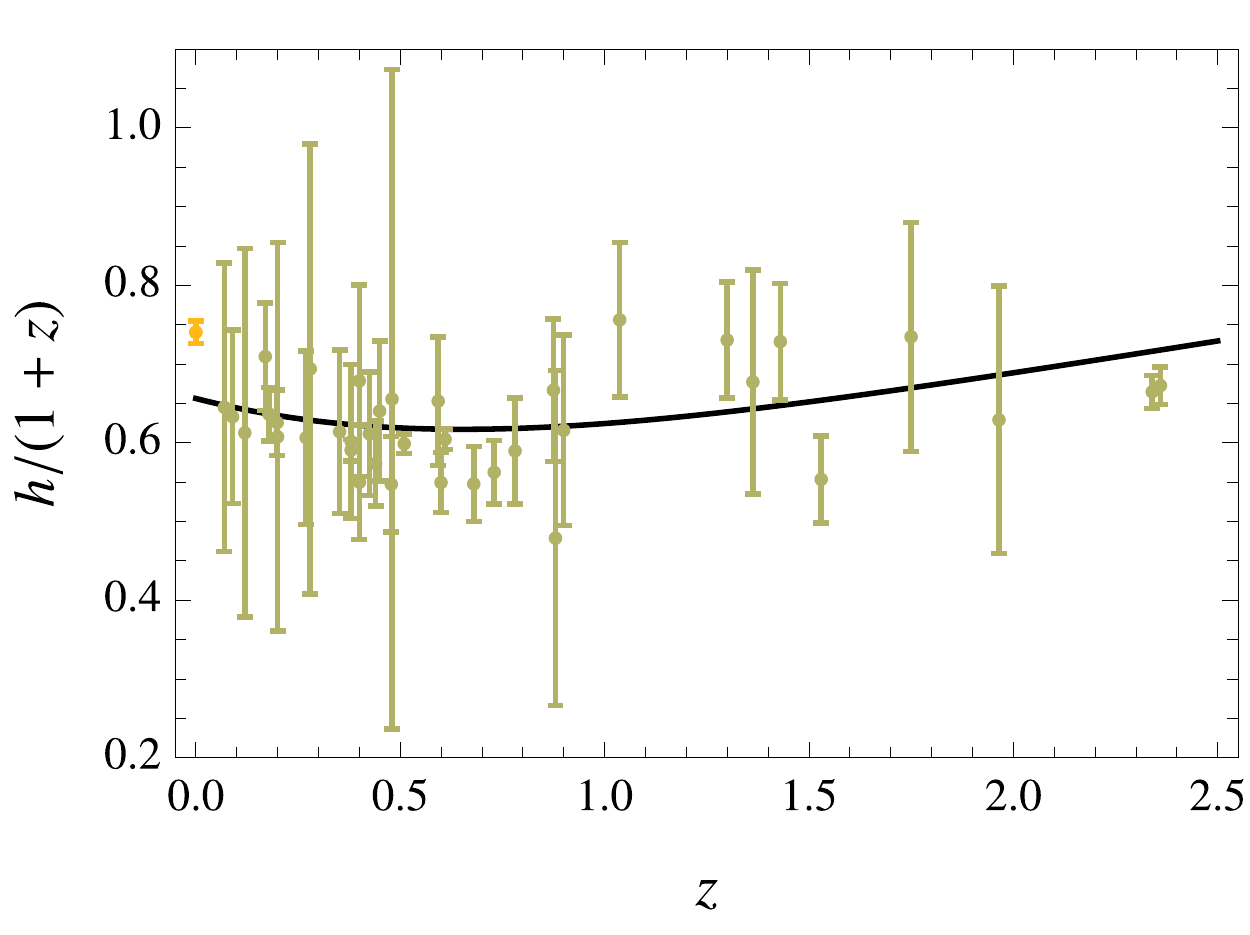}{0.99}
\end{minipage}
    \caption{Hubble expansion history for $z < 2.5$ considering $a=0.178$ (left) and
  $a=0$ (right). For comparison,
      we show the recent determination of $h_0$ from~\cite{Riess:2019cxk} together with a compilation~\cite{Farooq:2016zwm} of 38
measurements $h(z)$ in the range $0 \leq z
\leq 2.36$~\cite{Simon:2004tf,Sievers:2013ica,Moresco:2012jh,Zhang:2012mp,Font-Ribera:2013wce,Delubac:2014aqe,Moresco:2015cya,Moresco:2016mzx,Blake:2012pj,Alam:2016hwk}. These 38 $h(z)$ measurements are not
completely independent. For example, the 3 measurements taken
from~\cite{Blake:2012pj} are correlated with each other, and the 3
measurements of~\cite{Alam:2016hwk} are correlated too. In addition, in these and other cases, when BAO observations are used to measure $h(z)$, one has to apply a prior
on the radius of the sound horizon, $r_d =\int_{\rm z_d}^\infty c_s(z)dz/H(z)$, evaluated at the drag epoch $z_d$, shortly after
recombination, when photons and baryons decouple. This prior value of $r_d$ is usually derived using CMB observations. \label{fig:4}}
\end{figure}

The main results of the consistency investigation are encapsulated in Figs.~\ref{fig:1},
\ref{fig:2}, \ref{fig:3}, \ref{fig:q} and \ref{fig:4}, where we show the evolution
of: {\it (i)} $Y(u)$ and $Y'(u)$, {\it (ii)}~the various contributions
to the total energy density, {\it (iii)}~$w_y$, {\it (iv)}~the acceleration parameter $-q(u)=1+h'(u)/h(u)$, and {\it (v)} the
Hubble parameter.  The results shown in the left panels of these
figures are based on the best fit value of the AOV analysis (corresponding to $a =0.178$), whereas those
displayed in the right panels correspond to $a=0$. For $a=0$, we take $h_0
= 0.66$ and $w_{Y,0} = -0.70$; a choice justified below.  We can see in
Fig.~\ref{fig:2} how the $X$-$Y$ coupling depletes dark matter into
dark energy, yielding a larger $\Omega_{Y,0} = 0.704$ for $a = 0.178$
than for $a=0$ where $\Omega_{Y,0} = 0.673$. This is the so-called
``fading dark matter'' effect~\cite{Agrawal:2019dlm}, which tends to
favor larger values of $h_0$ when $a \neq 0$; namely, $h_0 = 0.69$ for
$a=0.178$, and $h_0= 0.66$ for $a=0$. The dark energy equation of state also
shows striking differences. As we can see in the left panel
Fig.~\ref{fig:3}, for $-10 \alt u \alt -2.$ the dark energy equation of
state $w_Y > 0$, so that the energy density redshifts faster than that
in $\Lambda$CDM~\cite{Agrawal:2019dlm}. For $a=0$, however, the dark
energy equation of state mimics that of a cosmological constant,
$w_Y = -1$, between $-10 \alt u \alt -2$. This translates into smaller
values of $w_{Y,0}$ for the decoupled system with $a=0$, and closer to
the $\Lambda$CDM prediction of $w_\Lambda =-1$.

\begin{figure}[tpb]
\begin{minipage}[t]{0.49\textwidth}
\postscript{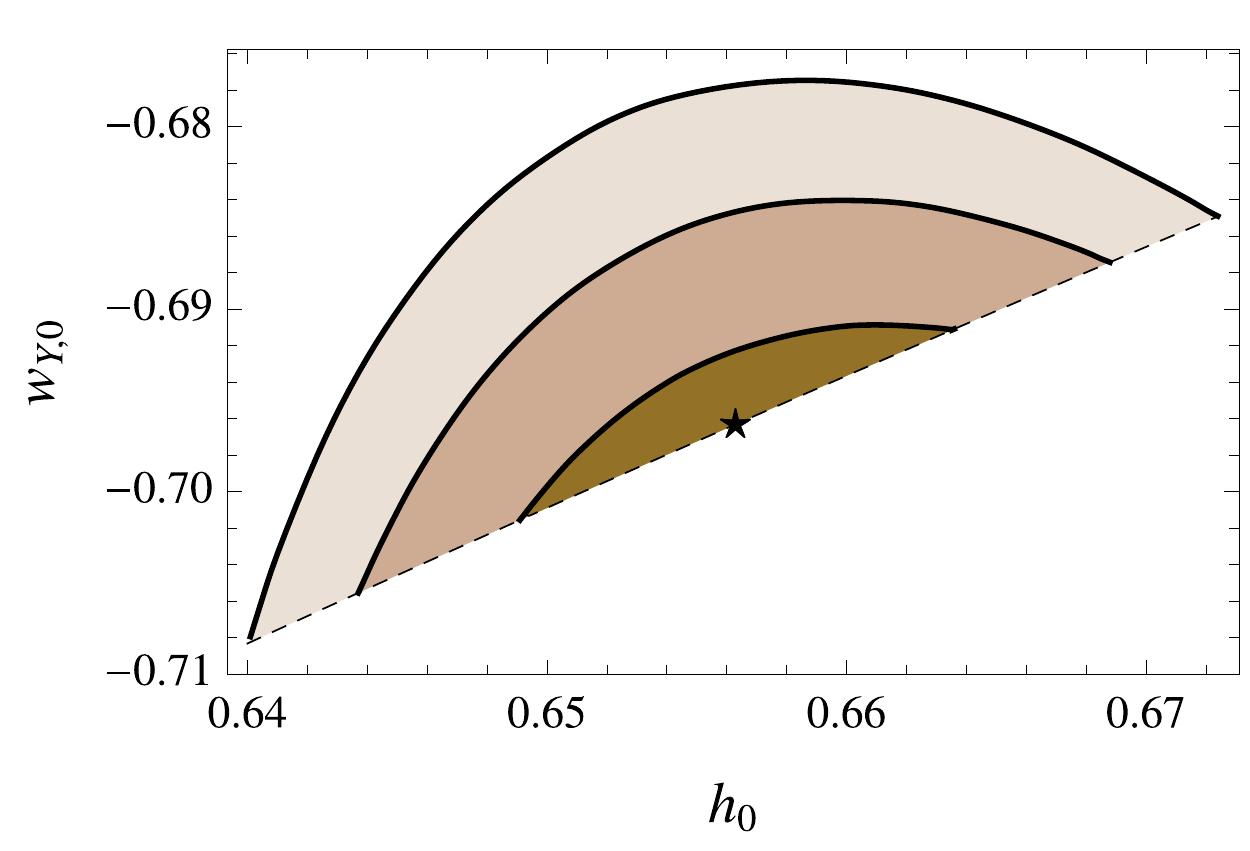}{0.99}
\end{minipage}
\begin{minipage}[t]{0.49\textwidth}
\postscript{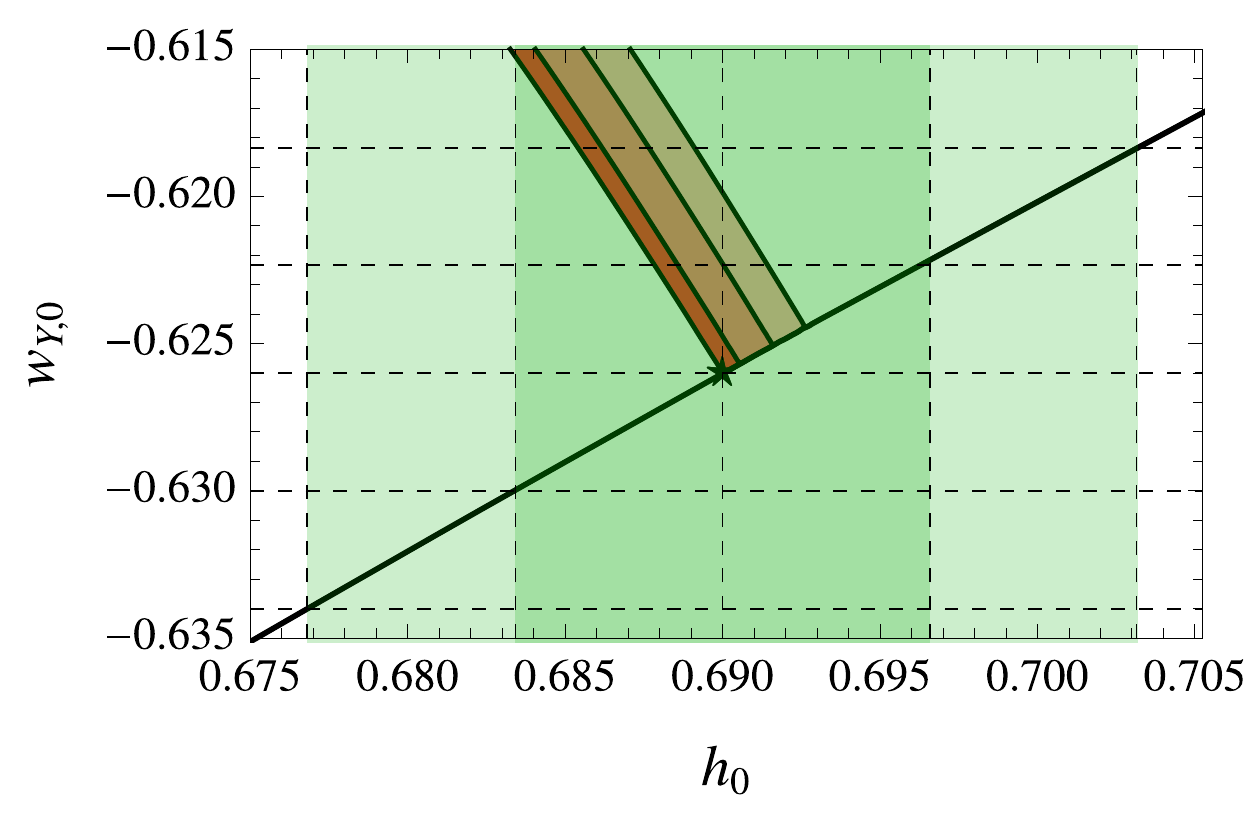}{0.99}
\end{minipage}
    \caption{{\it Left panel.} $1$, $3$ and $5\,\sigma$ probability contours in the
$(h_0,w_{Y,0})$ parameter space after performing a least squares fit of
the model to the Hubble parameter data. The minimum corresponds
to $a=0$. {\it Right panel.} Structure of the $(h_0,w_{Y,0})$ parameter space around the
point $(h_0,w_{Y,0})=(0.69,-0.626)$, indicated with a ``$\star$.'' The diagonal
line separates the regions with $\Omega_{X,0} /\Omega_{\mathrm{CDM},0}$
larger (above) and smaller (below) than $0.1$. The green bands indicate
the $1$ and $2\,\sigma$ confidence intervals for the value of $h_0$ as
determined by SH0ES. The colored contours show constant $\chi^2$ lines, after
the fit shown in the left panel. The first from the left shows the values as
likely as $(0.69,-0.626)$, and the other three show the values such
that $\Delta \chi^2$ is 1, 3 and 5, from left to right.
\label{fig:contours}}
  \end{figure}

In the left panel of Fig.~\ref{fig:contours} we show  $1$, $3$ and $5\,\sigma$ probability contours in the
$(h_0,w_{Y,0})$ parameter space after performing a least squares fit of
the model to the Hubble parameter data. The minimum, which corresponds
to $a=0$,  corroborates that quintessence models exacerbate the $H_0$ tension since the dark
energy density decreases in recent times~\cite{Agrawal:2019dlm}. As
can be seen in
the right panel of Fig.~\ref{fig:contours}, the best fit value of
$h_0$ in the OAV-study is consistent with determinations of $w_{Y,0}$
at $< 2\sigma$ level. We conclude that the set up introduced in Sec.~\ref{sec:2} has
the salient cosmological features of the AOV fading dark matter proposal.

\section{Hubble hullabaloo and D-brane string compactifications}
\label{sec:4}

 In this section we comment on additional
 phenomena that would influence the time evolution of the model
 parameters and may help solving the $H_0$ problem. It is common
 knowledge that D-brane string compactifications provide a collection
 of building block rules that can be used to build up the SM or
 something very close to
 it~\cite{Lust:2004ks,Blumenhagen:2005mu,Blumenhagen:2006ci}.  Gauge
 bosons of the brane stacks belong to ${\cal N} = 1$ vector multiplets
 together with the corresponding gauginos. At brane intersections
 chiral fermions belong to chiral multiplets denoted by their
 left-handed fermionic components $Q$, $L$, $U^c$, $D^c$, $E^c$,
 $N^c$, where the superscript $c$ stands for the charged conjugate in
 the familiar notation.

\begin{figure}[tb] 
    \postscript{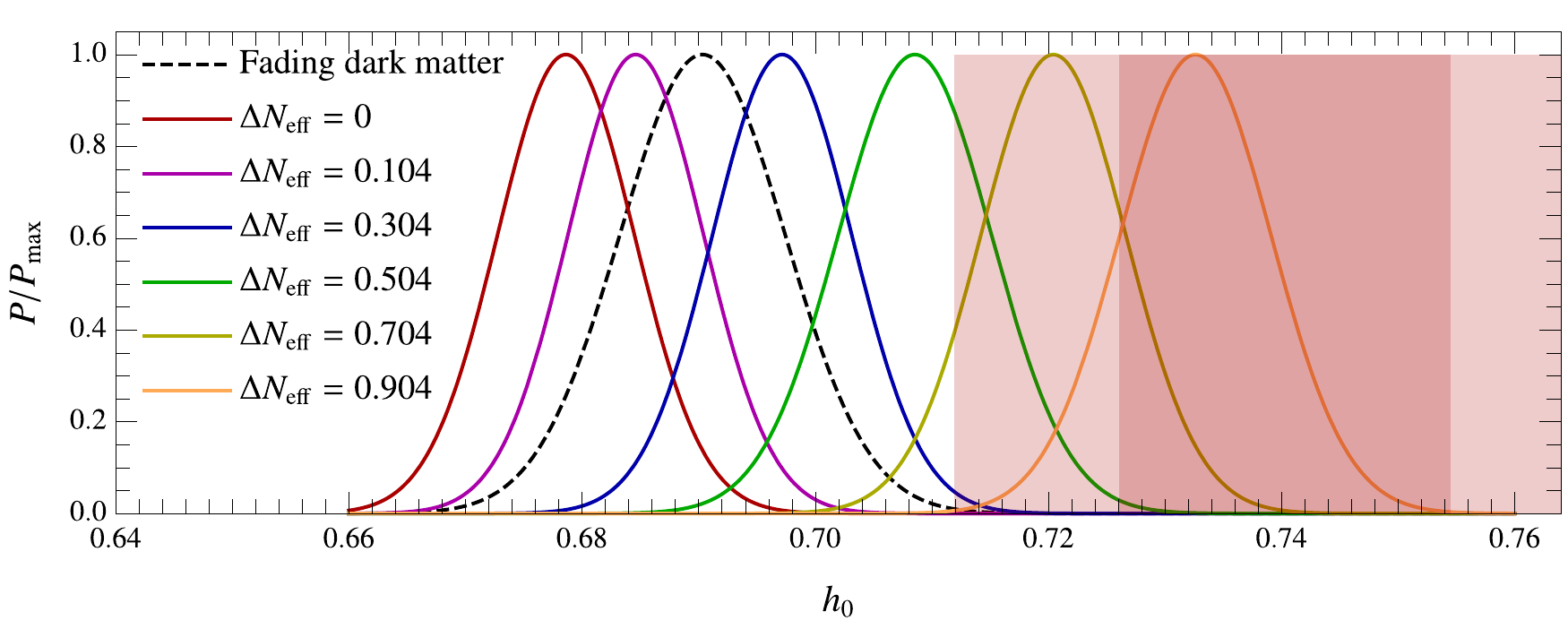}{0.99} 
    \caption{Rescaled posterior distributions of $h_0$ (due to marginalization over additional free parameters) with different
      choices of $N_{\rm eff}$ from the 7 parameter fir
      of~\cite{Vagnozzi:2019ezj}. The rescaled posterior distribution
      of $h_0$ for the AOV fit is indicated with the dashed curve~\cite{Agrawal:2019dlm}.The
      shaded areas indicate the $1\sigma$ and $2\sigma$ regions as
      determined by SH0ES~\cite{Riess:2019cxk}.
\label{fig:neff}}
\end{figure}
 
 For such $D$-brane constructs, superpotentials
 of the form $MN^cN^c$ or $SN^cN^c$ are precluded by the $U(1)_L$
 lepton and $U(1)_{I_R}$ isospin-right gauge invariances, where $M$ is
 a Majorana mass matrix in flavor space and $S$ is a gauge
 singlet. Because of this, there is no equivalent to the seesaw
 mechanism to generate the Weinberg term~\cite{Weinberg:1980bf} which
 gives rise to Majorana neutrinos.\footnote{However, it is possible
   that D-brane instantons can generate Majorana masses for these
   perturbatively forbidden
   operators~\cite{Blumenhagen:2006xt,Ibanez:2006da}.}  Neutrino
 masses could then depend upon the addition of 3 Dirac right-handed
 neutrinos. If we now adopt the phenomenological structure of D-brane
 models to describe the matter fields in the visible sector, then the
 model parameters of the cosmological set-up introduced herein could
 be (in principle) affected by the right-handed neutrinos, which would
 contribute to the total radiation energy density. For a decoupling
 temperature $\agt 1~{\rm TeV}$, we have
 $g_*(T_{\rm dec}) \agt 106.75$ and via (\ref{eq:Neff}) we find that
 the $\nu_R$ contribution to the non-SM relativistic energy density,
 $\Delta N_{\rm eff} \alt 0.14$, is well within the existing 95\% CL
 upper limit. On the other hand, if $\nu_R$'s decouple near the QCD
 phase transition, a D-brane-like description of the matter fields in
 our cosmological set-up can accommodate the larger value of
 $\Delta N_{\rm eff}$ derived using the helium abundance measurements
 of~\cite{Izotov:2014fga}, while providing interesting predictions for
 LHC
 searches~\cite{Anchordoqui:2011nh,Anchordoqui:2012wt,Anchordoqui:2012qu}.

 More concretely, in
 Fig.~\ref{fig:neff} we show the normalized posterior distributions of $h_0$ for
 different choices of $N_{\rm eff}$ from the 7 parameter fit of~\cite{Vagnozzi:2019ezj}. It is
 evident that the 95\% CL limit on $\Delta N_{\rm eff}$ from the
combination of CMB, BAO, and BBN observations~\cite{Aghanim:2018eyx} severely constrains a
solution of the $H_0$ problem in terms of additional relativistic
degrees of freedom. Consideration of the larger helium abundance
measured in~\cite{Izotov:2014fga}, with $\Delta N_{\rm eff} < 0.544$ at the 95\% CL still
precludes a full solution of the $H_0$ problem in terms of additional
light species at the CMB epoch. However, the combined effect
produced by fading dark matter and the extra relativistic
degrees of freedom at the CMB epoch appears to have the potential to
resolve the $H_0$ tension; see Fig.~\ref{fig:neff}. A comprehensive study of the parameter
space is beyond the scope of this paper and will be presented
elsewhere. Needless to say, the helium abundance
reported in~\cite{Izotov:2014fga} is in tension with Planck
observations, so this solution would
require a combination of two datasets which are in tension. On the one
hand,
the addition of extra relativistic degrees of freedom at the CMB epoch
can accommodate the local calibration of SNe luminosities
well out into the Hubble flow, avoiding the constraints of late times
dark energy transitions~\cite{Benevento:2020fev}. On the other hand,
we have noted that increasing the amount of radiation in the early
universe leads to a higher value of $S_8$. Solutions that could mitigate
this problem have been presented in~\cite{DiValentino:2018gcu}.

Future experiments, such as CMBPol (which is expected to reach a $2\sigma$
precision of $\Delta N_{\rm eff} = 0.09$~\cite{Galli:2010it}) and
eventually CMB-S4 (which is expected to reach a $2\sigma$ precision of
$\Delta N_{\rm eff} = 0.06$~\cite{Abazajian:2019eic}) will be able to
probe the contributions from $\varUpsilon$'s and $\nu_R$'s, providing
additional constraints on the (extended) string cosmological set-up proposed in this section.

\section{Conclusions}
\label{sec:5}

We have realized the Agrawal-Obied-Vafa swampland proposal of fading
dark matter for relaxing the $H_0$ tension~\cite{Agrawal:2019dlm} by
the model of Salam-Sezgin~\cite{Salam:1984cj} and its string
realization of Cveti\v c-Gibbons-Pope~\cite{Cvetic:2003xr}.  The model
is fairly simple, it describes a compactification from six to
four dimensions of a 6-dimensional SUGRA with a monopole background on a
2-sphere, allowing for time dependence of the 6-dimensional moduli
fields while assuming a 4-dimensional metric with a Robertson-Walker
form. In terms of linear combinations of the $S^2$ moduli field and
the 6-dimensional dilaton, the 4-dimensional effective potential
consists of a pure exponential function of a quintessence field $Y$ which is the 4-dimensional dilaton
and the field $X$ which determines the 4-dimensional gauge couplings
of NS5-branes. This avoids direct couplings of the dilaton to matter
suppressing extra forces competing with gravity. It turns out that $X$ is a source
of CDM, with a mass proportional to an exponential function of the
quintessence field. The asymptotic behavior of the Hubble parameter,
$h \approx \ln t$, leads to a conformally flat Robertson-Walker metric
for large times. The dS (vacuum) potential energy density is
characterized by an exponential behavior $V_Y \propto e^{\sqrt{2}
  Y}$. Asymptotically, this represents the crossover situation with
$w_Y = -1/3$, implying expansion at constant velocity with $Y$ varying logarithmically
$Y\approx -\ln t$~\cite{Antoniadis:1988aa}. 
The deviation from constant velocity expansion into a brief accelerated phase
encompassing the recent past ($z\alt 6$) makes the model
phenomenologically viable.

We have shown that this set up is well equipped to
reproduce the salient features of the AOV fading dark matter
proposal. For $a = 0.178$, the model features 
a tower of light states $X$ originating in the rolling of the $Y$
field. These $X$ particles constitute a portion of the CDM, and the way in
which their mass evolve over time demonstrates that the model may help reducing
(though not fully eliminate) the $H_0$ tension.

As a natural outgrowth of this work, we intend to study higher
dimensional SUGRAs, which also admit monopole-like
solutions~\cite{Salam:1989fm}. In some cases, however, there are no
compactifications to Minkowski vacuum~\cite{Salam:1984ft}. Of
particular interest is the gauged 8-dimensional SUGRA with matter
couplings~\cite{Salam:1985ns} where a solution of the form
Minkowski$_6 \times S^2$ is known to exist.  In addition, because the
Salam-Sezgin model has $N=(1,0)$ SUSY in 6 dimensions the $U(1)$
coupling is not fixed. In general it may be a combination of $e^\phi$
and $e^{-\phi}$ determined by chiral
anomalies~\cite{Seiberg:1996vs}. These may offer new possibilities for
models of the type discussed in this paper. However, we remind the
reader that late time dark energy transitions do not fully resolve the
true source of tension between the distance ladder and high redshift
observations~\cite{Benevento:2020fev} and therefore some additional
assumptions (like those discussed in Sec.~\ref{sec:4}) must be adopted
in order to solve the $H_0$ problem in the AOV-type string
backgrounds.

In summary, the string cosmological framework put forward in this
paper calls for new CMB
observations and stimulates the investigation of complex theoretical models of the swampland as possible solutions of the $H_0$ problem.

 \acknowledgments{We thank Eleonora Di Valentino for some valuable
   discussion. The work of L.A.A. and J.F.S. is supported by the
  by the U.S. National Science Foundation (NSF Grant PHY-1620661) and
  the National Aeronautics and Space Administration (NASA Grant
  80NSSC18K0464). The research of I.A. is funded in part by the
  “Institute Lagrange de Paris”, and in part by a CNRS PICS grant. The work of D.L. is
  supported by the Origins Excellence Cluster. The work of T.R.T is supported by
  NSF under Grant Number PHY–1913328.  Any opinions, findings, and
  conclusions or recommendations expressed in this material are those
  of the authors and do not necessarily reflect the views of the NSF
  or NASA.}

\section*{Appendix}

Since we have set to zero the fermionic terms in the background, the condition for the SUSY of the background is the vanishing of the
supersymmetric variations of the fermionic fields; namely,
  \begin{equation}
    \delta \chi = \frac{\kappa}{2} \left( \partial_M \phi \right)
    \Gamma^M  \epsilon + \frac{1}{12} e^{-\phi} G_{MNP} \Gamma^{MNP}
    \epsilon = 0 \,,
\label{app1}
  \end{equation}
\begin{equation}
 \delta \lambda = \frac{1}{2 \sqrt{2}} e^{-\phi/2} F_{MN} \Gamma^{MN}
 \epsilon - \frac{i}{\sqrt{2}} g e^{-\phi/2} \epsilon = 0 \,,
\label{app2}
\end{equation}
and
\begin{equation}
  \delta \psi_M = \frac{1}{\kappa} \mathscr{D}_M \epsilon +
  \frac{1}{24} e^{-\phi} G_{PQR} \Gamma^{PQR} \Gamma_M \epsilon = 0 \,,
\label{app3}
\end{equation}
for the
axino,  the dilatino, and the gravitino; respectively~\cite{Salam:1984cj}. Here,
$\Gamma^{PQR} = \Gamma^{[P} \Gamma^Q \Gamma^{R]}$ is the fully
antisymmetric product of three $\Gamma$-matrices of the 6-dimensional
Clifford algebra. The covariant derivative of the gravitino,
\begin{equation}
 \mathscr{D}_M \psi_N = \left(\partial_M + \frac{1}{4} \omega_{MAB}
   \Gamma^{AB} - i g A_M\right) \psi_N \,, 
\end{equation}
is given in terms of the torsion-free spin connection
$\omega_M^{AB}$. (The Christoffel connection is not needed because of
the contraction with the antisymmetric gamma-matrix.) Using the
vielbein $e_A^M$, we have
\begin{equation}
  \omega_M^{AB} = 2 e^{N[A} \partial_{[M}{e_{N]}}^{B]} - e^{N[A} e^{B]P}
  e_{MC} \partial_N e_P^C \, .
  \end{equation}
In familiar notation: $\Gamma_\mu =
\gamma_\mu \times \sigma^1$, $\Gamma_5 = \gamma_5 \times \sigma^1$,
$\Gamma_6 = \mathds{1} \times \sigma^2$, $\{\Gamma_M,\Gamma_N\} = 2
\eta_{MN}$, $\gamma_5^2 = \mathds{1}$ and so $\Gamma_{56} = \gamma_5
\times i \sigma^3$ and $\Gamma_7 = \Gamma_0 \Gamma_1 \cdots \Gamma_6 =
\mathds{1} \times \sigma^3$~\cite{RandjbarDaemi:1982hi}.

With this in mind, the non-zero components of the spin connection 
are found to be $\omega_{\hat \imath}^{i 0} = e^h (\dot f + \dot
h)$, $\omega_{\hat 5}^{0 5} = r_c \dot f$,
$\omega_{\hat 6}^{0 6} = r_c \dot f \sin
\vartheta$, $\omega_{\hat 6}^{5  6} = \cos
\vartheta$, where we adopted the also familiar notation of carets on the
curved indices (which are lowered or raised with the spacetime metric $g_{MN}$) to
distinguish them from the flat indices (that are lowered or raised
with the flat Minkowski metric $\eta_{AB}$), so that $g^{MN} =
\eta^{AB} e_A^M e_B^N$; lowercase latin indices are used for the 3
spatial components of $M_4$, and run from 1 to 3. The contraction $F_{MN}
\Gamma^{MN}$ in (\ref{app2}) takes the form
\begin{equation}
 F_{MN} \ \Gamma^{MN} = 2 F_{\hat 5 \hat 6} \ \Gamma^{\hat 5 \hat 6} =
 2 b \
 \sin \vartheta \ e_{5}^{\hat 5} \ e_{6}^{\hat 6} \ \Gamma^{56} = 2
 \frac{b}{r_c^2} \ e^{-2f} \ \Gamma_{56} \, .
\label{app6}
\end{equation}
Substituting (\ref{app6}) into (\ref{app2}) we obtain
\begin{equation}
 \frac{1}{\sqrt{2}} \ e^{-\phi/2} \ \left[b \ e^{-2f} \ \Gamma_{56} \ \epsilon -
     i g \ e^{\phi} \ \epsilon \right] =0 \, .
\label{app7}
 \end{equation}
Remarkably, the field equations fixed the monopole charged to be
$\pm 1$, and lead to the condition~\cite{Salam:1984cj}
 \begin{equation} 
 \Gamma_{56} \epsilon = \pm i \epsilon \, .
\label{app8}
\end{equation}
 Using (\ref{app8}) we rewrite (\ref{app7}) as
 \begin{equation}
  e^{2f+ \phi} = \pm \frac{b}{g}  \, .
\label{app9}
\end{equation}
 In a similar fashion, $\delta \psi_{\hat 0} = 0$ leads to
\begin{equation}
 \partial_0
 \epsilon =0 \, ,
\label{app10}
\end{equation}
from which we conclude that $\epsilon$ is not a function of $t$.  The
condition $\delta \psi_{\hat \imath} = 0$ yields
\begin{equation}
  \partial_i \epsilon + \frac{1}{2} \ \omega_{\hat \imath}^{i0} \
  \Gamma_{i0} \
  \epsilon =  \partial_i \epsilon + \frac{1}{2} \ e^h \ (\dot f + \dot
  h) \
  \Gamma_{i0} \ \epsilon =0 \, ,
\label{app11}
\end{equation} 
  and $\delta \psi_{\hat 5} = 0$ gives
  \begin{equation}
    \partial_5 \epsilon + \frac{1}{2} \ \omega_{\hat 5}^{50} \
    \Gamma_{50} \
    \epsilon - ig \ A_{\hat 5} \ \epsilon = \partial_5 \epsilon +
    \frac{1}{2} \ r_c \
    \dot f  \ \Gamma_{50} \ \epsilon = 0 \, .
\label{app12}
  \end{equation}
 Next, $\delta \psi_{\hat 6} = 0$, leads to
    \begin{equation}
      \partial_6 \epsilon + \frac{1}{2} \ \omega_{\hat 6}^{60} \
      \Gamma_{60} \ \epsilon + \frac{1}{2} \ \omega_{\hat 6}^{56} \
      \Gamma_{56} \ \epsilon - ig \ A_6 \ \epsilon = 0 \,,
      \label{app13}
    \end{equation}
    which translates into
    \begin{equation}
      \partial_6 \epsilon + \frac{1}{2} r_c \dot f \ \sin \vartheta \
      \Gamma_{60} \ \epsilon = 0
\label{app14}
    \end{equation}
    and
    \begin{equation}
      \frac{1}{2} \ \cos \vartheta \ \Gamma_{56} \ \epsilon - ig \ b \
      \cos \vartheta \ \epsilon = 0 \, .
\label{app15}
    \end{equation}
Substituting the relation $g = \sqrt{\xi}/2$ into
(\ref{app15}) while imposing (\ref{app8})  we obtain the
contraint $b^2 \xi = 1$. Finally, the variation of $\delta \chi$
implies
\begin{equation}
  \frac{\kappa}{2}  \ \partial_0 \phi \ \Gamma_{\hat 0} \ \epsilon = 0 \, ,
\label{app16}  
\end{equation}
which sets $\dot \phi =0$.

The constraints from imposing
the SUSY background can be summarized as follows: the relation (\ref{app16})
demands $\phi$ to be a constant and when this condition is combined
with (\ref{app9}) we see that $f$ must also be a constant. Because $f$
is a constant, we can immediately see by
inspection of (\ref{app10}), (\ref{app12}), and (\ref{app14}) that
$\epsilon$ is independent of both $t$ and the coordinates of the
compact space $\vartheta$ and $\varphi$. Likewise, we rewrite  (\ref{app11}) as
\begin{equation}
  \partial_i \epsilon + \frac{1}{2} e^h \ \dot h \ \Gamma_{i0} \ \epsilon = 0 \, .
\label{app17}
\end{equation}
The temporal dependence of (\ref{app17}) then becomes
\begin{equation}
  e^h \ \dot h = \varkappa_1 \,,
\label{app18}
\end{equation}
and so the scale factor for a SUSY background is found to be
\begin{equation}
  e^h = \varkappa_1 t + \varkappa_2 \,,
\label{app19}
\end{equation}
  with $\varkappa_1$ and $\varkappa_2$ constants.

\end{document}